\documentclass[letterpaper]{article} % DO NOT CHANGE THIS
\usepackage{aaai25}  % DO NOT CHANGE THIS
\usepackage{times}  % DO NOT CHANGE THIS
\usepackage{helvet}  % DO NOT CHANGE THIS
\usepackage{courier}  % DO NOT CHANGE THIS
\usepackage[hyphens]{url}  % DO NOT CHANGE THIS
\usepackage{graphicx} % DO NOT CHANGE THIS
\usepackage{natbib}  % DO NOT CHANGE THIS AND DO NOT ADD ANY OPTIONS TO IT
\usepackage{caption} % DO NOT CHANGE THIS AND DO NOT ADD ANY OPTIONS TO IT
\usepackage{algorithm}
\usepackage{algorithmic}

\usepackage{multirow}
\usepackage{soul}
\usepackage{amsmath,amsfonts,amsthm,bm}
\usepackage{newfloat}
\usepackage{listings}
\DeclareCaptionStyle{ruled}{labelfont=normalfont,labelsep=colon,strut=off} % DO NOT CHANGE THIS
\floatstyle{ruled}
\newfloat{listing}{tb}{lst}{}
\floatname{listing}{Listing}
\newtheoremstyle{propositionstyle}%
  {3pt}% space above
  {3pt}% space below
  {\itshape} % body font
  {}% indent
  {\bfseries}% theorem head font
  {:}% punctuation after theorem head
  {0.5em}% space after theorem head
  {}% theorem head spec
\theoremstyle{propositionstyle}
\newtheorem{proposition}{Proposition}

\newtheoremstyle{assumptionstyle}%
  {3pt}% space above
  {3pt}% space below
  {}% body font
  {}% indent
  {\bfseries}% assumption head font
  {:}% punctuation after assumption head
  {0.5em}% space after assumption head
  {}% assumption head spec

\theoremstyle{assumptionstyle}
\newtheorem{assumption}{Assumption}

\title{Enhancing Adversarial Transferability with Adversarial Weight Tuning}
\author{
   Jiahao Chen\textsuperscript{\rm 1}\equalcontrib, Zhou Feng\textsuperscript{\rm 1}\equalcontrib, Rui Zeng\textsuperscript{\rm 1}, Yuwen Pu\textsuperscript{\rm 1}, Chunyi Zhou\textsuperscript{\rm 1}\thanks{Corresponding author.}, \\ Yi Jiang\textsuperscript{\rm 1}, Yuyou Gan\textsuperscript{\rm 1}, Jinbao Li\textsuperscript{\rm 2}\textsuperscript{\rm 3} and Shouling Ji\textsuperscript{\rm 1}\\
}
\affiliations{
    \textsuperscript{\rm 1} College of Computer Science and Technology, Zhejiang University\\
    \textsuperscript{\rm 2} Shandong Artificial Intelligence Institute\\
    \textsuperscript{\rm 3} School of Mathematics and Statistics, Qilu University of Technology\\
}

\usepackage{bibentry}
\begin{document}

\maketitle

\begin{abstract}
Deep neural networks (DNNs) are vulnerable to adversarial examples (AEs) that mislead the model while appearing benign to human observers. A critical concern is the transferability of AEs, which enables black-box attacks without direct access to the target model. However, many previous attacks have failed to explain the intrinsic mechanism of adversarial transferability. In this paper, we rethink the property of transferable AEs and reformalize the formulation of transferability. Building on insights from this mechanism, we analyze the generalization of AEs across models with different architectures and prove that we can find a local perturbation to mitigate the gap between surrogate and target models. We further establish the inner connections between model smoothness and flat local maxima, both of which contribute to the transferability of AEs. Further, we propose a new adversarial attack algorithm, \textbf{A}dversarial \textbf{W}eight \textbf{T}uning (AWT), which adaptively adjusts the parameters of the surrogate model using generated AEs to optimize the flat local maxima and model smoothness simultaneously, without the need for extra data. AWT is a data-free tuning method that combines gradient-based and model-based attack methods to enhance the transferability of AEs. Extensive experiments on a variety of models with different architectures on ImageNet demonstrate that AWT yields superior performance over other attacks, with an average increase of nearly 5\% and 10\% attack success rates on CNN-based and Transformer-based models, respectively, compared to state-of-the-art attacks.

\end{abstract}


\section{Introduction}

In recent years, deep neural networks (DNNs) have achieved remarkable success across a wide range of applications, from computer vision~\cite{he2016deep} and natural language processing~\cite{radford2019language} to speech recognition~\cite{povey2011kaldi} and autonomous driving~\cite{caesar2020nuscenes}. The proliferation of large-scale models~\cite{rombach2022high}, empowered by advancements in computational resources and the availability of vast datasets, has further accelerated the adoption of DNNs in real-world scenarios. Despite these achievements, DNNs have been shown to be vulnerable to adversarial examples (AEs)~\cite{goodfellow2014explaining,madry2017towards}, which are intentionally crafted to mislead the model while appearing benign to human observers.

\begin{figure}[t]
    \centering
    \includegraphics[width=0.8\linewidth]{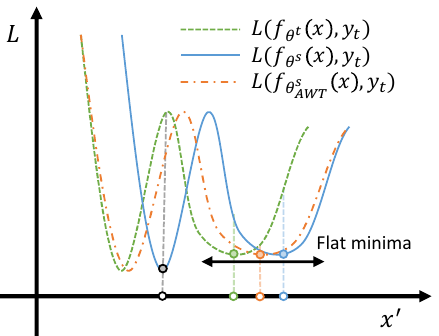}
    \caption{The motivation of the proposed AWT that achieves flat local minima via adjusting the surrogate model’s parameters $\theta^{s}$ to enable more transferable AEs. This plot illustrates that we can achieve low $\nabla_{x^{\prime}}\ell(x^{\prime},y;\theta^{s})$ through the input and parameter space. The colored dots represent the generated AEs with different loss value on the target model $\theta^{t}$.}
    \label{fig:illustration}
\end{figure}

From a practical perspective, one particularly concerning aspect of AEs is their ability to transfer between different models~\cite{ma2023transferable,zhang2023transferable}, allowing attackers to craft effective perturbations without direct access to the target model's architecture or parameters~\cite{dong2018boosting,wang2021boosting, wang2021enhancing}. Recently, researchers have proposed a variety of transferable attack methods aiming to maximize the success rate of these attacks and avoid over-fitting to the surrogate models~\cite{dong2018boosting, dong2019evading}. These methods leverage different strategies, such as input transformation by applying data augmentation~\cite{xie2019improving}, model ensemble that exploits the outputs of multiple models in loss or logits~\cite{liu2016delving} and gradient rectification with momentum~\cite{dong2018boosting,lin2019nesterov}, to generate transferable AEs across diverse architectures of the surrogate model. Although these methods have presented better performance, there are no clear explanations for the factors behind transferability.

To explain what factors have resulted in the different transferability of different models, Wu et al.~\cite{wu2018understanding} explored how adversarial transferability relies on the model-specific factors empirically and found that local non-smoothness of the loss surface harms the transferability of generated AEs. Next, Yang et al.~\cite{yang2021trs} first theoretically linked two key factors, i.e., model smoothness and gradient similarity, with the lower bound of AEs transferability. With this intuition, researchers have proposed to finetune the attacker's surrogate model to enhance its smoothness~\cite{zhu2022toward,wu2024lrs,zhang2024does}, since AEs generated on smoothed surrogate models have shown better transferability. Nevertheless, these methods all assume that attackers can obtain extra data with a similar distribution of the target models' training data, and time-consuming tuning on these large-scale datasets introduces computation overhead as well. Different from these model-based attacks, recent works~\cite{ge2023boosting,qiu2024enhancing} can achieve state-of-the-art (SOTA) performance on transferability by optimizing the flat local maxima (minima) of AEs, as illustrated (from black to blue dot) in Figure~\ref{fig:illustration}.

While these methods have made significant strides in improving the transferability, \textbf{limitaions} remain. (1) Many of the existing methods rely on empirical evaluations with different models, lacking a unified and representative metric to measure transferability across various architectures. (2) Moreover, model-based methods require access to extra data (the same distribution as that of the target model), which may not always be feasible in real-world scenarios. (3) Model-based methods only optimize smoother landscapes on benign samples, leading to suboptimal performance since AEs are often generated iteratively. (4) Both gradient and model-based methods fail to explain why flat local maxima or model smoothness boosts the generalization of AEs across different models. Therefore, it is necessary to formalize the definition of transferable AEs and find the connection with the aforementioned factors, which will offer more insights into understanding adversarial transferability.

To this end, we rethink the property of transferable AEs and develop a novel formulation from the perspective of generalization to measure transferability. Building upon the insights from this formulation, we first analyze the generalization of AEs across models with different architectures and prove that we can always find a local perturbation to mitigate the gap between surrogate and target models. Further, with these insights, we observe inner connections between model smoothness and flat local maxima, which both contribute to the transferability of AEs. Based on this, we then propose a new adversarial attack algorithm, \textbf{A}dversarial \textbf{W}eight \textbf{T}uning (AWT), which adaptively adjusts the parameters of the surrogate model using generated AEs to optimize the flat local maxima and model smoothness simultaneously, without the need for extra data. As shown in Figure~\ref{fig:illustration}, better transferability (orange dot) can be achieved through input and parameter space. We evaluate AWT on a variety of models with traditional experiments and the metric derived from our proposed formulation. Our contributions are as follows:
\begin{itemize}
\item We theoretically analyze the transferability of AEs and reveal the inherent factors of the surrogate model and AEs that account for the transferability of AEs, which have been neglected by previous works. 
\item Based on the analysis above, a data-free tuning method AWT is developed by adaptively adjusting the parameters of the surrogate model while optimizing adversarial perturbations with better transferability. To the best of our knowledge, AWT is the first method that combines gradient-based and model-based attack methods to enhance the transferability of AEs.
\item Further, with our formulation as a measurement of adversarial transferability across models. We conduct comprehensive experiments for evaluation with CNN-based and Tranformer-based models on ImageNet and demonstrate that, AWT yields superior attack success rates and metric values over eight SOTA transferable attacks. 
\end{itemize}

\section{Related Work}
\subsection{Gradient-based Adversarial Attack}

Gradient-based methods are among the most common approaches to generating AEs and improving their transferability. These methods basically involve crafting perturbations based on the gradients of the loss function with respect to the input. One of the earliest and most influential gradient-based methods is the Fast Gradient Sign Method (FGSM)~\cite{goodfellow2014explaining}, which adds a small perturbation to the input in the direction of the sign of the gradient. Subsequent improvements, such as the Iterative FGSM~\cite{kurakin2016adversarial} and the Momentum-based FGSM~\cite{dong2018boosting,lin2019nesterov}, iteratively refine the perturbation to achieve higher success rates and better transferability. Later, researchers proposed methods to improve the transferability of AEs by leveraging additional gradient information. For example, the Virtual Momentum Iterative FGSM (VMI)~\cite{wang2021enhancing} and the Ensemble Momentum Iterative FGSM (EMI)~\cite{wang2021boosting} use virtual and ensemble gradients, respectively, to enhance the transferability of AEs. More recently, the connection between flat local maxima and generalization~\cite{keskar2016large,foret2020sharpness} have inspired researchers to generate AEs at a flat local region of the loss function, making them more likely to generalize across models. For instance, Qin et al.~\cite{qin2022boosting} seek to minimize the loss of AEs at a region with low loss value by adding reverse perturbations before calculating gradients. Ge et al. and Qiu et al.~\cite{ge2023boosting,qiu2024enhancing} introduce approximated Hessian-vector products and aggregate the gradients of randomly sampled data around AEs.

\subsection{Model-based Adversarial Attack}
There are also many works focusing on enhancing the smoothness of the surrogate model since Wu et al.~\cite{wu2018understanding} found that local non-smoothness of the loss surface harms the transferability of AEs, leading to methods that finetune the surrogate model to be smoother~\cite{zhu2022toward,wu2024lrs}.
These model-based methods generate transferable AEs by modifying the surrogate model itself, which often involve pre-processing steps that modify the model's architecture~\cite{li2020learning} or training loss function. For example, Distribution-Relevant Attack (DRA)~\cite{zhu2022toward} achieves smoother by minimizing the Hessian-vector products of the surrogate model.

\subsection{Others}
In addition to gradient-based and model-based methods, researchers have explored other techniques to improve the transferability of AEs~\cite{Wang22mm,Wang21cvpr,Wang24sp,Wang24ijcv}.
One approach is to apply input transformations, such as data augmentation, to the surrogate model during training. Techniques like Diverse Input (DIM)~\cite{xie2019improving} and Translation-invariant method (TIM)~\cite{dong2019evading} have been shown to improve the transferability of AEs by making the AEs more robust to data transformations. Another line of research focuses on optimizing AEs with the aggregated gradients from the ensemble of surrogate models~\cite{liu2016delving,gubri2022lgv}. With augmentation of surrogate models, these methods aim to create AEs that are able to fool a wider range of target models with significantly different architectures.

\section{Methodology}
\subsection{Preliminaries}
Formally, take an image classification task as an example, let $\mathcal{F}: \mathcal{X}\rightarrow\mathcal{Y}$ denotes the set of mapping function with parameters, where input space $\mathcal{X}=\mathbb{R}^{B\times H \times W \times C}$ and output space $\mathcal{Y}=\mathbb{R}^{B\times K}$ ($B$ and $K$ means batch size and the number of the classes, respectively). For a dataset $\mathcal{D}=\{(x_1,y_1),(x_2,y_2),...(x_i,y_i),...(x_N,y_N)\}$, where $x_i\in\mathcal{X}$ and $y_i\in\mathcal{Y}$. Under white-box settings, attackers can generate AEs $x^{\prime}$ by solving the maximization problem with $f_{\theta}$:

\begin{equation}
    \label{eq:eq1}
    \max_{x^{\prime}} \ell(x^{\prime},y;\theta), \quad\mathbf{s.t.} \|x-x^{\prime}\|_{p}\leq\epsilon,
\end{equation}
where $f_{\theta}\in\mathcal{F}$, $\ell(\cdot,\cdot;\cdot)$ is the loss function, and $\|x-x^{\prime}\|_{p}\leq\epsilon$ restricts the $L_p$ norm of the discrepancy between $x$ and $x^{\prime}$, where we adopt $p=\infty$ in this paper. 

In black-box scenarios, different from Eq.(\ref{eq:eq1}), the attacker aims to generate AEs with high generalization (transferability) across models with different architectures and parameters. To achieve this, the attacker needs to solve the following problem: 
\begin{equation}
\label{eq:eq2}
\max_{x^{\prime}}\mathbb{E}_{f_{\theta_i}\in\mathcal{F}}\left[\ell(x^{\prime},y;\theta_i)\right],
\end{equation}
where $f_{\theta_i}$ denotes the target models trained on $\mathcal{D}$ from $\mathcal{F}$. Since the architecture and parameters of the target model $\theta^{t}$ are out of reach, $x^{\prime}$ can only generated on surrogate model $\theta^{s}$ held by attackers. Given that the optimization problem in Eq.(\ref{eq:eq2}) involves unknown models, we first simplify this by assuming that our target models are trained on $\mathcal{D}$ with the same architecture but different parameters. Therefore, we can simplify Eq.(\ref{eq:eq2}) as follows:

\begin{equation}
\label{eq:eq4}
\max_{x^{\prime}}\mathbb{E}_{\eta_{i}\in\Omega}\left[\ell(x^{\prime},y;\theta^{s}\oplus\eta_{i})\right],
\end{equation}
where $\theta^{s}\oplus\eta_{i}$ means that we can add different perturbation $\eta$ on parameter space $\theta^{s}$ to stand for different target models with the same architecture but different parameters. $\Omega$ represents the set of parameter perturbations. 

\subsection{Enhance Transferability with Local (Min)Maxima}
With the preliminaries above, we aim to find the inner connection between transferability and corresponding factors like flat local maxima and model smoothness in this subsection. Based this connection, we then solve the problem in Eq.(\ref{eq:eq4}) with local maxima (minima) from the perspective of both input and parameter space to achieve transferability.

\textbf{Achieving Flat Local Maxima from Input Space.} From the perspective of generalization, to achieve Eq.(\ref{eq:eq2}), recent works~\cite{ge2023boosting,qiu2024enhancing} proposed to find a flat local maxima of AEs to achieve better transferability:
\begin{equation}
    \label{eq:eq3}
    \max_{x^{\prime}} \ell(x^{\prime},y;\theta^{s}) - \lambda
    \|\nabla_{x^{\prime}}\ell(x^{\prime},y;\theta^{s})\|_{2}.
\end{equation}
However, the mechanism behind this factor remains to be discussed and explored. Here, we present a theoretical analysis on this factor. 
\begin{proposition}\label{pro1}
Given a model $f_{\theta}$ with parameters $\theta$, and a perturbation $\eta$ such that $\|\eta\|_{p} \leq \kappa$ ($\kappa \rightarrow 0$), we want to prove that $\forall x\in\mathcal{D}$ and $\gamma \rightarrow 0$, there exists an perturbation $\delta$ such that:
\begin{equation}
\|f_{\theta+\eta}(x) - f_{\theta}(x+\delta)\|_{p} \leq \gamma.
\end{equation}
\end{proposition}
Proposition \ref{pro1} states that models with different parameters can be approximated by introducing perturbations in the input space, which suggests that we can bridge the gap between target and surrogate models using input perturbations. Therefore, we can further approximate Eq.(\ref{eq:eq4}) as follows:
\begin{equation}
\label{eq:eq6}
\max_{x^{\prime}}\mathbb{E}_{\delta_{i}\in\hat{\Omega}}\left[\ell(x^{\prime}+\delta_{i},y;\theta^{s})\right],
\end{equation}
where $\delta_{i}\in\hat{\Omega}=\left\{ \delta \,\middle|\, \|f_{\theta + \eta}(x) - f_{\theta}(x + \delta)\|_p \leq \gamma, \eta\in\Omega \right\}$ represents the perturbation set in the input space to eliminate the differences in parameter space. For the generation of AEs, we can transform Eq.(\ref{eq:eq6}) into a bi-level optimization problem by maximizing the lower bound of the expectation item in Eq.(\ref{eq:eq6}):
\begin{equation}
\label{eq:eq7}
\max_{x^{\prime}}\min_{\delta}\left[\ell(x^{\prime}+\delta,y;\theta^{s})\right].
\end{equation}
Subsequently, we further approximate the first-order Taylor expansion of the loss function $\ell(\cdot)$ at the point $x^{\prime}$:
\begin{equation}
    \ell(x^{\prime}+\delta,y;\theta^{s}) = \ell(x^{\prime},y;\theta^{s}) + \delta^{T} \cdot \nabla_{x^{\prime}}\ell(x^{\prime},y;\theta^{s}) + \mathcal{O}(\Vert\delta\Vert_2^2).
\end{equation}
Please note that solving the inner optimization of Eq.(\ref{eq:eq7}) can be approximated by $-\alpha\nabla_{x^{\prime}}\ell(x^{\prime},y;\theta^{s})$. This approximation leads to an equation nearly equivalent to Eq.(\ref{eq:eq3}), which have actually been proven in NCS~\cite{qiu2024enhancing}. Consequently, we can relate the flat local maxima $\|\nabla_{x^{\prime}}\ell(x^{\prime},y;\theta^{s})\|_{2}$ to the transferability of AEs across models with the same architecture but different parameters, as outlined in Eq.(\ref{eq:eq6}).

However, SOTA attacks such as NCS~\cite{qiu2024enhancing} and PGN~\cite{ge2023boosting} achieve flat local maxima of $x^{\prime}$ solely through optimization in the input space. In contrast, we propose achieving this by adjusting the parameters of the surrogate models. Unlike model-based tuning methods, AWT leverages the intermediate AEs during the optimization process. Since using $\|\nabla_{x^{\prime}}\ell(x^{\prime},y;\theta^{s})\|_{2}$ as loss for model optimization is non-trivial, we use generated $x^{\prime}_t$ to achieve this:
\begin{equation}
\label{eq:eq9}
\begin{aligned}
    \ell(x^{\prime}_{t},y;\theta^{s}) &=  (x^{\prime}_{t}-x^{\prime}_{t-1})^{T} \cdot \nabla_{x^{\prime}_{t-1}}\ell(x^{\prime}_{t-1},y;\theta^{s})\\
    &+\ell(x^{\prime}_{t-1},y;\theta^{s}) + \mathcal{O}(\Vert\delta\Vert_2^2),
\end{aligned}
\end{equation}
where $x^{\prime}_{t}$ denotes the generated AEs in the $t$-th iteration, and $(x^{\prime}_{t+1}-x^{\prime}_{t})^{T} \cdot \nabla_{x^{\prime}}\ell(x^{\prime},y;\theta^{s})$ approximates $\|\nabla_{x^{\prime}}\ell(x^{\prime},y;\theta^{s})\|_{2}$. Consequently, we can also achieve flat local maxima by minimizing $\ell(x^{\prime}_{t},y;\theta^{s})$. Additionally, we add the term $\ell(x,y;\theta^{s})$ to maintain the stability of the surrogate model. Thus, the overall loss function $\mathcal{L}_{AWT}$ for model tuning at the iteration $t$ is given by:
\begin{equation}
    \mathcal{L}_{AWT} = \ell(x^{\prime}_{t},y;\theta^{s}) + \ell(x,y;\theta^{s}).
\end{equation}

\begin{figure}[t]
    \centering
    \includegraphics[width=0.9\linewidth]{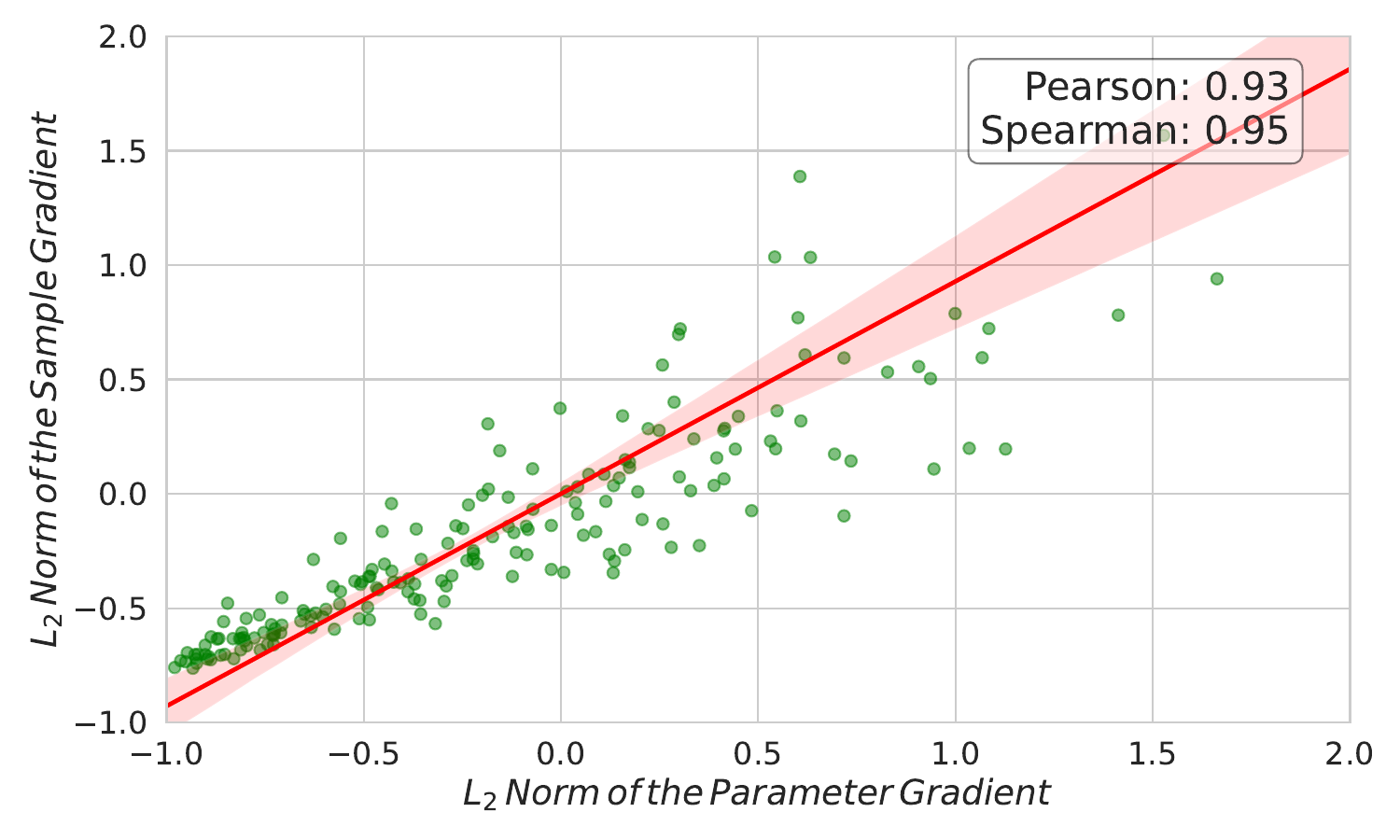}
    \caption{Correlation between the $L_2$ norms of sample and parameter gradient (with normalization: $g=\frac{g-\mu}{\sigma}$). }
    \label{fig:grad_norm}
\end{figure}

\begin{algorithm}[t]
\caption{Adversarial Weight Tuning (AWT) attack}
\label{alg:awt}
\begin{algorithmic}[1]
\REQUIRE Benign input $x$ with ground-truth label $y$; the loss function $\ell$ with surrogate model $f_{\theta^{s}}$; the magnitude of perturbation $\epsilon$; the maximum number of iterations $T$; the decay factor $\mu$; the balanced coefficient $\omega$; the step size of parameters: $\beta$, $lr$; the number of randomly sampled examples $N$.
\ENSURE An adversarial example $x^{\prime}$.
\STATE $g_0=0$, $x_0^{\prime}=x$, $\theta_{temp}=\theta^{s}$
\FOR{$t=0,1,\cdots,T-1$}
    \STATE Set $\bar{g}=0$
    \STATE Compute the perturbed $\hat{\theta}^{s}$ with $\theta^{s} + \beta\cdot\nabla_{\theta^{s}}\mathcal{L}_{AWT}$
    \STATE Update surrogate $\theta^{s}$ with $ \theta^{s} - lr\cdot\nabla_{\hat{\theta}^{s}}\mathcal{L}_{AWT}$
    \STATE // \textit{Adjust the parameters of $\theta^{s}$ to obtain surrogate model with better transferability}.
    \FOR{$i=0,1,\cdots,N-1$}
        \STATE Randomly sample an example $x^* \sim \mathcal{N}(x_t^{\prime},\zeta)$
        \STATE Calculate the gradient at $x^*$: $g^\prime=\nabla_{x^*} \ell(x^*,y;\theta^{s})$
        \STATE Compute the predicted point by $x^* = x^* - \alpha \cdot \frac{g^\prime}{\|g^\prime\|_1}$
        \STATE Calculate the gradient at $x^*$: $g^*=\nabla_{x^*} \ell(x^*,y;\theta^{s})$
        \STATE Update gradient by $\bar{g} = \bar{g} + \frac{1}{N}[(1-\omega) \cdot g^\prime + \omega \cdot g^*]$
    \ENDFOR
    \STATE $g_{t+1} = \mu \cdot g_t + \frac{\bar{g}}{\|\bar{g}\|_1}$
    \STATE Update $x_{t+1}^{\prime}$ via $x_{t+1}^{\prime} = \Pi_{\mathcal{B}_\epsilon(x)}[x_t^{\prime} + \alpha \cdot \text{sign}(g_{t+1})]$
\ENDFOR
\STATE $\theta=\theta_{temp}$
\RETURN $x^{\prime} = x_T^{\prime}$
\end{algorithmic}
\end{algorithm}

\textbf{Achieving Flat Local Minima from Parameter Space.} In addition to the flat local maxima of AEs, Proposition~\ref{pro1} suggests that model smoothness also contributes to adversarial transferability. 
\begin{proposition}
\label{pro2}
    For a model $f_{\theta}$ parameterized by $\theta$ and a loss function $\ell(x, y; \theta)$, there is a positive correlation between the $L_2$ norm of the input gradient $\|\nabla_{x}\ell(x,y;\theta)\|_2$ and that of the parameter gradient $\|\nabla_{\theta}\ell(x,y;\theta)\|_2$.
\end{proposition}
Proposition~\ref{pro2} indicates that the $L_2$ norms of the input gradient and the parameter gradient are positively correlated. To empirically validate this correlation, we visualize $\|\nabla_{x}\ell(x,y;\theta)\|_2$ and $\|\nabla_{\theta}\ell(x,y;\theta)\|_2$ for each sample in Figure~\ref{fig:grad_norm}, from which we can observe a strong positive correlation between the $L_2$ norm of the sample gradient and that of the parameter gradient, with Pearson and Spearman correlation coefficients of 0.93 and 0.95, respectively. Leveraging this correlation, we can instead minimize $\|\nabla_{\theta^s}\ell(x^{\prime},y;\theta)\|_2$. However, directly incorporating the gradient norm as a loss term introduces significant computational overhead due to the necessity of computing second-order derivatives. To address this challenge, we propose reformulating the problem as a bi-level optimization:
\begin{equation}
    \min_{\theta^{s}}\max_{\|\eta\|_2\leq\rho} \mathcal{L}_{AWT}(x_{t}^{\prime},x,y;\theta^{s}\oplus\eta),
\end{equation}
where $\eta$ denotes the $l_2$ norm bound of parameter perturbations. In this framework, the inner optimization problem implicitly minimizes the gradient norm by seeking the most challenging perturbations under which the model should remain robust. Simultaneously, the outer optimization adjusts the parameters $\theta^{s}$ to minimize the worst-case loss, thereby enhancing model smoothness. This approach preserves the benefits of gradient norm minimization while avoiding prohibitive computational costs.

For practical implementation, we incorporate previous gradient-based techniques~\cite{qiu2024enhancing,ge2023boosting} (flat local maxima of input space) to achieve flat local maxima in both input and parameter spaces, thereby improving the generalization of AEs across different models. The complete algorithm for generating AEs is detailed in Algorithm~\ref{alg:awt}, which also ensures a low $\|\nabla_{\theta^s}\ell(x^{\prime},y;\theta)\|_2$ by optimizing samples around $x^{\prime}$.

\begin{table*}[htbp]
\centering
\tabcolsep=0.15cm
\renewcommand{\arraystretch}{1.05}
\begin{tabular}{|c|c|ccccccccc|}
\hline
Model & Attack & ViT-B/16 & ViT-B/32 & Swin-T & Swin-S & ConViT & LeViT & CAiT-S & DeiT-B & PiT-B \\ \hline
\multirow{9}{*}{RN50} & MI & 11.7±0.3 & 11.6±0.2 & 25.1±0.6 & 20.6±0.5 & 11.5±0.3 & 17.8±0.2 & 12.3±0.4 & 10.9±0.5 & 15.5±0.3 \\
 & NI & 11.3±0.3 & 11.3±0.3 & 25.2±0.1 & 20.9±0.4 & 11.2±0.2 & 17.7±0.2 & 11.6±0.4 & 10.7±0.2 & 14.5±0.1 \\
 & VMI & 28.6±0.2 & 21.6±0.3 & 42.5±0.3 & 38.1±0.3 & 27.4±0.4 & 36.0±0.4 & 29.4±0.4 & 28.5±0.9 & 33.3±0.3 \\
 & VNI & 23.0±0.7 & 17.4±0.4 & 40.8±0.5 & 35.4±0.3 & 23.0±0.2 & 33.5±0.6 & 24.4±0.4 & 23.6±0.3 & 29.3±0.4 \\
 & EMI & 19.3±0.2 & 15.6±0.3 & 39.5±0.6 & 32.8±0.1 & 19.8±0.4 & 30.6±0.4 & 20.9±0.5 & 19.2±0.4 & 25.1±0.3 \\
 & RAP & 10.8±0.2 & 10.9±0.2 & 23.9±0.1 & 19.9±0.6 & 10.6±0.1 & 16.5±0.2 & 11.4±0.1 & 9.9±0.2 & 14.6±0.4 \\
 & PGN & \ul{49.2±0.4} & \ul{40.8±0.2} & \ul{66.0±0.1} & 59.2±0.2 & 46.8±0.4 & \ul{60.4±0.6} & 50.9±0.2 & 49.5±0.3 & 53.0±0.2 \\
 & NCS & 49.1±0.3 & 39.7±0.3 & 63.7±0.1 & \ul{59.4±0.2} & \ul{48.0±0.2} & 58.8±0.5 & \ul{51.5±0.2} & \ul{50.6±0.3} & \ul{54.1±0.5} \\
 & AWT & \textbf{59.0±0.3} & \textbf{46.6±0.5} & \textbf{72.5±0.3} & \textbf{67.6±0.4} & \textbf{56.6±0.4} & \textbf{69.0±0.3} & \textbf{61.3±0.2} & \textbf{60.0±0.4} & \textbf{62.7±0.5} \\ \hline
\multirow{9}{*}{RN101} & MI & 18.5±0.4 & 14.4±0.2 & 31.6±0.6 & 28.1±0.4 & 18.7±0.5 & 25.8±0.9 & 19.2±0.1 & 18.3±0.3 & 23.3±0.4 \\
 & NI & 17.3±0.3 & 13.5±0.3 & 31.2±0.6 & 27.8±0.9 & 17.8±0.5 & 25.5±0.2 & 18.1±0.4 & 16.9±0.2 & 22.0±0.2 \\
 & VMI & 36.0±0.2 & 26.2±0.2 & 48.8±0.3 & 45.3±0.5 & 35.8±0.6 & 44.3±0.7 & 37.9±0.4 & 37.0±0.6 & 41.0±0.9 \\
 & VNI & 31.0±0.8 & 21.9±0.7 & 47.5±0.4 & 42.2±1.1 & 31.0±0.6 & 42.1±0.9 & 33.5±0.5 & 32.2±0.7 & 37.8±1.1 \\
 & EMI & 30.4±1.0 & 21.3±0.2 & 51.5±0.5 & 44.8±0.5 & 30.8±0.4 & 44.8±0.6 & 33.5±0.5 & 31.8±0.4 & 37.7±0.5 \\
 & RAP & 16.4±0.3 & 13.6±0.1 & 29.4±0.3 & 25.6±0.3 & 16.7±0.2 & 23.8±0.1 & 17.3±0.3 & 15.6±0.2 & 20.9±0.2 \\
 & PGN & \ul{63.0±0.5} & \ul{51.9±0.3} & \ul{73.4±0.3} & \ul{69.4±0.2} & \ul{60.2±0.5} & \ul{70.7±0.4} & \ul{65.9±0.3} & \ul{63.5±0.1} & \ul{65.8±0.1} \\
 & NCS & 58.5±0.1 & 48.3±0.4 & 69.7±0.2 & 66.3±0.2 & 57.5±0.3 & 66.0±0.3 & 62.5±0.2 & 59.6±0.1 & 62.6±0.1 \\
 & AWT & \textbf{68.0±0.4} & \textbf{55.1±0.4} & \textbf{76.8±0.2} & \textbf{73.8±0.2} & \textbf{65.9±0.3} & \textbf{75.3±0.6} & \textbf{71.3±0.1} & \textbf{68.9±0.5} & \textbf{70.1±0.2} \\ \hline
\end{tabular}
\caption{The attack success rates (\%±std, over 10 random runs) of nine gradient-based attacks on nine normally trained Transformer-based models. The best and second results are \textbf{bold} and \ul{underline}, respectively.}
\label{tab:transformer_attack}
\end{table*}

\section{Evaluation}
\subsection{Experimental Settings}    
\textbf{Dataset.} Evaluation is conducted on the ImageNet-compatible dataset, which is widely utilized in prior work~\cite{qin2022boosting,ge2023boosting,qiu2024enhancing}. This dataset comprises 1,000 images which we resized with a resolution of 224$\times$224$\times$3 pixels, along with their corresponding ground-truth labels.

\textbf{Models.} To demonstrate the efficacy of our methods, we evaluate attack performance on several popular pre-trained models including \textbf{CNN-based} and \textbf{Transformer-based} models. For CNN models, we consider: ResNet-18 (RN18), ResNet-50 (RN50), and ResNet-101(RN101)~\cite{he2016deep}, Inception-v3 (Inc-v3)~\cite{szegedy2016rethinking}, Inception-v4 (Inc-v4), InceptionResNet-v2 (IncRes-v2)~\cite{szegedy2017inception}, VGG-13, VGG-16~\cite{simonyan2014very} and DenseNet-121 (Dense121)~\cite{huang2017densely}. Additionally, we consider Transformer-based models, including ViT-Base/16 (ViT-B/16), ViT-Base/32 (ViT-B/32)~\cite{dosovitskiy2020image}, PiT-B~\cite{heo2021rethinking}, Swin-T, Swin-S~\cite{liu2021swin}, ConViT~\cite{d2021convit}, LeViT~\cite{graham2021levit}, CAiT-S~\cite{touvron2021going}, DeiT-B~\cite{touvron2021training}

\textbf{Compared Methods.} As for baselines, we employ eight popular gradient-based iterative adversarial attacks: MI~\cite{dong2018boosting}, Nesterov Accelerated FGSM (NI)~\cite{lin2019nesterov}, VMI~\cite{wang2021enhancing}, EMI~\cite{wang2021boosting}, and Robust Adversarial Perturbations (RAP)~\cite{qin2022boosting}, Probabilistic Gradient Noise (PGN)~\cite{ge2023boosting} and Neighborhood Conditional Sampling (NCS)~\cite{qiu2024enhancing}.

\textbf{Hyper-parameters.} Following the previous work~\cite{ge2023boosting,qin2022boosting,qiu2024enhancing}, we set the maximum perturbation $\epsilon = 16.0 / 255$, the number of iterations $T = 10$, and the step size $\alpha = 1.6 $. For MI and NI, the decay factor $\mu = 1.0$. For VMI, we set the number of sampled examples $N = 20$ and the upper bound of the neighborhood size $\beta = 1.5 \times \epsilon$. For EMI, we set $N = 11$, the sampling interval bound $\eta = 7$, and use linear sampling. For the RAP attack, we set $\alpha = 2.0 / 255$, $K = 400$, the inner iteration number $T = 10$, the late-start $K_{LS} = 100$, and the size of neighborhoods $\epsilon_n = 16.0 / 255$. For PGN, NCS and AWT, we set $N = 20$, the balanced coefficient $\delta = 0.5$, and the upper bound $\zeta = 3.0 \times \epsilon$. For AWT alone, we set $\beta=0.005$ and $lr=0.002$ for surrogate models used for evaluation.

\textbf{Note.} \textit{Due to the extensive amount of evaluation results, we do not include all the data in the main body of the paper. The complete set of results is available in the appendix.}

\subsection{Measurement of Adversarial Transferability}
With Equation~\ref{eq:eq4}, we wonder if it is possible to derive a metric for the measurement of transferability, which can also serve as a metric to measure the transferability of AEs. Specifically, we propose quantifying the difference between the benign model and the perturbed one using AEs. Here, we define the transferability score $T$ as follows:
\begin{equation}
\label{eq:metric}
T(\mathcal{A},\theta^{s}) = \mathbb{E}_{\substack{x\in\mathcal{D} \\ \eta_{i}\in\Omega}}\left[\|f_{\theta^{s}\oplus\eta_{i}}(\mathcal{A}(x)) - f_{\theta^{s}}(\mathcal{A}(x))\|_2\right],
\end{equation}
where $\mathcal{A}$ denotes the attack method. The transferability score $T$ measures the expected difference in loss between the surrogate model and the perturbed one when evaluated on AEs. A lower value of $T$ indicates a higher transferability, meaning that the AEs generated on the $\theta^{s}$ are more likely to fool the target models. Considering that obtaining $\Omega$ is not trivial, we assume that the $l_2$ norm of the perturbation in $\Omega$ is bounded. Specifically, we sample $\theta\in\mathcal{B}_{\varepsilon}(\theta)$, where $\mathcal{B}_{\varepsilon}(\theta)$ denotes the $\varepsilon$-ball around $\theta$, to approximate the target models. The proposed metric can be used to evaluate the effectiveness of various attack strategies in generating AEs that are transferable across different models. For the hyper-parameters of the metric in Eq.(\ref{eq:metric}), we sample 10 perturbations around the parameters of the surrogate models with the distribution $\mathcal{N}(\theta^{s},\varepsilon)$ and $\varepsilon\in\{0.001,0.01,0.1\}$.

\subsection{Experimental Results}
 
\textbf{Evaluation on Transformer-based Models.} 
With the rise of large models in various domains, transformer architecture models have become increasingly prevalent due to their superior performance in handling complex tasks. The success of these models has led to a surge in their adoption across a wide range of applications. Therefore, we mainly focus on evaluating the effectiveness of the AWT method in generating AEs with high transferability on Transformer-based models. Table~\ref{tab:transformer_attack} presents the attack success rates (ASR) of nine gradient-based attacks on nine pretrained Transformer-based models. The results are averaged over 10 random runs, and the standard deviations are reported alongside the mean values. As shown in Table~\ref{tab:transformer_attack}, among the existing methods, PGN and NCS perform particularly well, often achieving the second-highest ASR. However, AWT consistently outperforms these methods, demonstrating the superiority of our method in generating AEs with high transferability. For instance, on the ViT-B/16 model (target), AWT achieves an ASR of 59.0\%, surpassing the second-best method, PGN, which has an ASR of 49.2\%. Similarly, on the RN101 (surrogate), AWT achieves an ASR of 68.0\%, outperforming PGN, which has an ASR of 63.0\%. 

\begin{table}[t]
\centering
\tabcolsep=0.1cm
\renewcommand{\arraystretch}{1}
\begin{tabular}{|c|ccccc|}
\hline
Model & Attack & Inc-v4 & Inc-v3 & Dense121 & VGG-16 \\ \hline
\multirow{9}{*}{RN50} & MI & 26.7±0.2 & 32.8±0.8 & 39.1±0.7 & 40.4±0.5 \\
 & NI & 26.3±0.5 & 32.8±0.2 & 40.5±0.5 & 41.8±0.3 \\
 & VMI & 46.9±0.5 & 49.1±0.3 & 58.1±0.7 & 56.5±0.4 \\
 & VNI & 45.5±0.3 & 47.6±0.6 & 60.6±1.0 & 58.2±0.8 \\
 & EMI & 43.1±0.7 & 48.1±0.9 & 64.3±0.7 & 63.9±0.8 \\
 & RAP & 25.1±0.2 & 30.8±0.7 & 36.4±0.6 & 38.9±0.5 \\
 & PGN & \ul{73.2±0.3} & \ul{75.3±0.3} & \ul{82.5±0.2} & \ul{80.2±0.2} \\
 & NCS & 67.8±0.1 & 68.5±0.2 & 76.5±0.2 & 73.9±0.1 \\
 & AWT & \textbf{77.0±0.3} & \textbf{79.1±0.3} & \textbf{86.7±0.2} & \textbf{84.6±0.3} \\ \hline
\multirow{9}{*}{RN101} & MI & 36.6±0.3 & 40.5±0.6 & 49.2±0.4 & 46.9±0.4 \\
 & NI & 36.7±0.7 & 40.2±0.4 & 50.0±0.9 & 48.0±0.9 \\
 & VMI & 55.2±0.3 & 57.2±0.4 & 66.0±1.0 & 60.0±0.2 \\
 & VNI & 55.4±0.5 & 56.2±1.0 & 69.1±0.3 & 61.7±0.3 \\
 & EMI & 59.6±0.8 & 61.3±0.4 & 76.4±0.3 & 70.1±0.4 \\
 & RAP & 34.2±0.4 & 38.4±0.9 & 46.0±1.0 & 44.9±0.7 \\
 & PGN & \ul{82.1±0.1} & \ul{81.0±0.2} & \ul{86.1±0.1} & \ul{82.1±0.3} \\
 & NCS & 75.0±0.4 & 74.2±0.2 & 80.4±0.1 & 76.5±0.3 \\
 & AWT & \textbf{82.4±0.2} & \textbf{82.0±0.2} & \textbf{87.4±0.3} & \textbf{84.9±0.2} \\ \hline
\end{tabular}
\caption{The attack success rates (\%±std, over 10 random runs) of nine gradient-based attacks on four normally trained CNN-based models. The best and second results are \textbf{bold} and \ul{underline}, respectively.}
\label{tab:cnn_attack}
\end{table}

\textbf{Evaluation on CNN-based Models.} Building on the strong performance of the AWT on Transformer-based models, we now evaluate its effectiveness on CNN-based models. Table~\ref{tab:cnn_attack} presents the ASR of nine gradient-based attacks on four pretrained CNN-based models, including Inc-v4, Inc-v3, DN121, and VGG16. As shown in Table~\ref{tab:cnn_attack}, AWT consistently outperforms other attacks, achieving the highest ASRs across all CNN-based models. For instance, on the Inc-v4 model, AWT achieves an ASR of 77.0\%, surpassing the second-best method, PGN, which has an ASR of 73.2\%. Similarly, on the RN101 model, AWT achieves an ASR of 82.4\%, outperforming PGN, which has an ASR of 82.1\%. The superior performance of AWT on CNN-based models further supports its effectiveness in generating AEs with high transferability. These results are consistent with those observed on Transformer-based models, indicating that AWT is a robust method capable of enhancing the transferability of AEs across different model architectures.

\begin{table*}[htb]
\centering
\tabcolsep=0.1cm
\renewcommand{\arraystretch}{1}
\begin{tabular}{|c|ccccccccc|}
\hline
Attack & Inc-v4 & Inc-Res-v2 & Inc-v3 & DN121 & VGG16 & ViT-B/16 & Swin-T & LeViT & DeiT-B \\ \hline
DIM & 43.3 & 38.1 & 46.7 & 55.1 & 52.3 & 23.7 & 37.1 & 31.4 & 24.8 \\
DIM+AWT & \textbf{48.5} & \textbf{42.2} & \textbf{50.6} & \textbf{60.4} & \textbf{59.7} & \textbf{27.9} & \textbf{43.4} & \textbf{34.5} & \textbf{28.4} \\ \cline{1-10} 
TIM & 31.5 & 24.2 & 34.0 & 41.1 & 43.7 & 18.0 & 27.2 & 21.3 & 13.8 \\
TIM+AWT & \textbf{39.4} & \textbf{31.2} & \textbf{41.3} & \textbf{50.4} & \textbf{51.1} & \textbf{21.2} & \textbf{33.6} & \textbf{26.0} & \textbf{17.9} \\ \cline{1-10} 
Admix & 49.7 & 42.1 & 52.1 & 65.1 & 62.2 & 27.0 & 44.6 & 36.8 & 27.2 \\
Admix+AWT & \textbf{61.3} & \textbf{52.9} & \textbf{61.6} & \textbf{75.7} & \textbf{72.3} & \textbf{32.6} & \textbf{54.2} & \textbf{46.6} & \textbf{32.2} \\ \cline{1-10} 
SIM & 39.3 & 31.1 & 43.2 & 53.9 & 51.6 & 18.6 & 36.0 & 25.8 & 18.4 \\
SIM+AWT & \textbf{58.2} & \textbf{49.1} & \textbf{58.5} & \textbf{71.7} & \textbf{66.5} & \textbf{31.6} & \textbf{51.0} & \textbf{42.0} & \textbf{31.4} \\ \hline
\end{tabular}
\caption{The attack success rates (\%) of AWT method, when it is integrated with DIM, TIM, SIM and Admix, respectively. The AEs are generated on RN50. The best and second results are \textbf{bold}.}
\label{tab:ablaion_attack}
\end{table*}

\textbf{Integrated with Other Attacks.} Having demonstrated the effectiveness of the AWT (only model tuning) method on both CNN- and Transformer-based models, we now investigate its impact when integrated with other attacks which also follows the approach taken in previous works~\cite{qiu2024enhancing}. AWT can further enhance the adversarial transferability of other attacks without any conflict or overlap, effectively complementing each other. Specifically, we integrate AWT with input transformation-based attacks, such as DIM~\cite{xie2019improving}, TIM~\cite{dong2019evading}, SIM~\cite{lin2019nesterov}, and Admix~\cite{wang2021admix}, and evaluate the ASR on a variety of target models. Table~\ref{tab:ablaion_attack} presents the ASR of these attacks when integrated with AWT, using RN50 and RN101 as the surrogate models. The best results are highlighted in bold. As shown in Table~\ref{tab:ablaion_attack}, integrating AWT with existing attack methods consistently improves the ASRs across all target models. For instance, when integrated with DIM, AWT increases the ASR from 43.3\% to 48.5\% on the Inc-v4 model, and from 55.2\% to 58.1\% on the Inc-v3 model when using RN50 as the surrogate model. Similarly, when integrated with TIM, AWT improves the ASR from 36.9\% to 42.9\% on Inc-v3 when using RN101 as the surrogate model. The improvements are substantial across all attack methods. For example, when integrated with Admix, AWT increases the ASR from 62.3\% to 66.2\% on Inc-v3 when using RN50 as the surrogate model. The integration of AWT with other attacks demonstrates its versatility and ability to enhance the transferability of AEs.

\begin{figure}[t]
    \centering
    \includegraphics[width=0.9\linewidth,height=150pt]{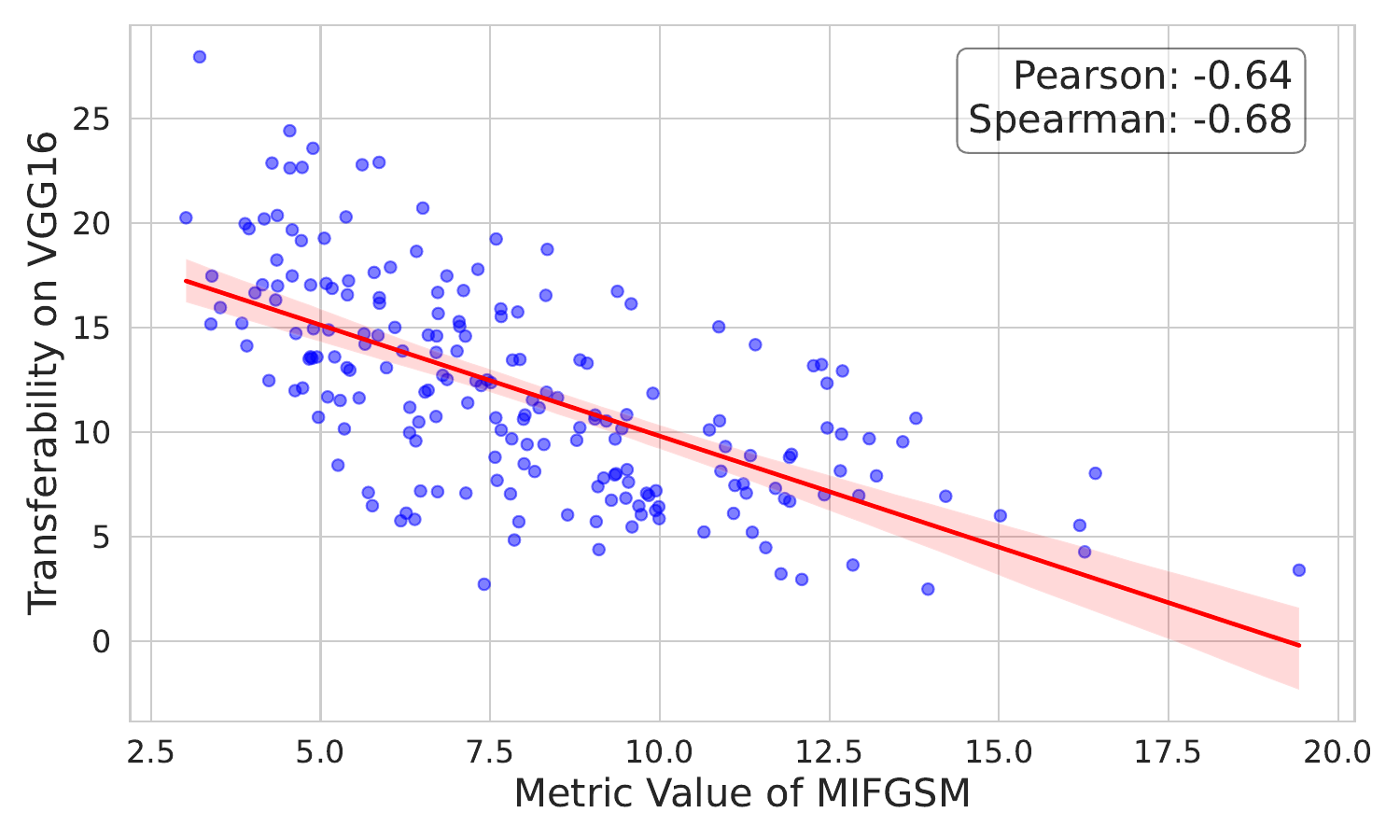}
    \caption{Relationship between the proposed metric and experimental transferability on VGG16 (target model). Each dot represents an adversarial sample. Note that AEs are generated on RN50 and the amplitude $\varepsilon$ of the parameter perturbation is set to $0.05$. }
    \label{fig:transfer}
\end{figure}

\begin{table*}[htb]
\centering
\tabcolsep=0.15cm
\renewcommand{\arraystretch}{1}
\begin{tabular}{|c|c|ccccccccc|}
\hline
Model & $\varepsilon$ & MI & NI & VMI & VNI & EMI & RAP & PGN & NCS & AWT \\ \hline
\multirow{3}{*}{RN50}
 & 0.001 & 0.193 & 0.192 & 0.190 & 0.192 & 0.252 & 0.221 & 0.224 & 0.197 & \textbf{0.184} \\
 & 0.010 & 1.751 & 1.816 & 1.687 & 1.867 & 2.128 & 2.136 & 1.978 & 1.646 & \textbf{1.642} \\
 & 0.100 & 15.934 & 16.240 & 18.397 & 19.288 & 19.043 & 16.911 & 16.643 & 16.351& \textbf{15.852} \\ \hline
\multirow{3}{*}{RN101} & 0.001 & 0.154 & 0.145 & 0.146 & 0.147 & 0.161 & 0.165 & 0.159  & 0.144 & \textbf{0.142}\\
 & 0.010 & 1.500 & 1.394 & 1.374 & 1.360 & 1.583 & 1.471 & 1.512  & 1.382& \textbf{1.283} \\
 & 0.100 & 15.095 & 13.017 & 14.125 & 15.719 & 15.506 & 13.570 & 14.247 & 14.480& \textbf{12.121} \\ \hline
\end{tabular}
\caption{The metric value of different attacks with different amplitude of parameter perturbation on surrogate models. Note that $\varepsilon$ denotes the amplitude of the perturbation mentioned in Eq.\ref{eq:metric} and the lower value means higher transferability.}
\label{tab:metric}
\end{table*}

\textbf{Transferability Measurement.} To better understand the relationship between our proposed formulation (metric) (Eq.(\ref{eq:metric})) and the actual transferability, we first analyze the correlation between the metric values and the transferability obtained on target model (VGG16). Figure~\ref{fig:transfer} depicts the relationship between the proposed metric and the transferability on the VGG16 model with 100 generated samples. For experimental transferability (y-axis), we use $\|f_{\theta^{t}}^{y}(x)-f_{\theta^{t}}^{y}(x^{\prime})\|$ empirically, which represents how adversarial perturbations influence the target model from the confidence (without softmax) of the benign label $y$. The x-axis represents the calculated metric value, while the y-axis shows the corresponding transferability achieved on the VGG16 model. As shown in Figure~\ref{fig:transfer}, there exists a strong negative correlation between the metric value and the transferability on VGG16. The Pearson coefficient of $-0.64$ and the Spearman coefficient of $-0.68$ indicate a significant inverse relationship between the two variables. This observation suggests that higher metric values correspond to lower transferability, providing evidence that our metric can capture the characteristics of transferable AEs to a certain degree. 

Further, with the evidence above, we use our proposed metric to evaluate nine attacks under different parameter perturbations with RN50 and RN101 and the results are shown in Table~\ref{tab:metric}. We can observe that AWT can achieve the lowest metric value among all attacks, suggesting that it generates AEs with better transferability. This finding aligns with the results reported earlier in Table~\ref{tab:transformer_attack} and Table~\ref{tab:cnn_attack}.
However, we can also notice there are discrepancies between experimental results and our metric. For instance, PGN and NCS tend to perform better transferability than other attacks but they sometimes have larger metric values than attacks than VMI and VNI. This can be attributed to the insufficient sampling of our metric, leading to significant errors. Therefore, it is crucial to acknowledge that the transferability of AEs in real-world settings can be significantly more complex due to the diversity of model architectures. Given these complexities, our proposed metric, while effective in certain contexts, may not fully capture the intricacies of transferability in practice. (\textbf{Limitations}) (1) The metric assumes a certain level of similarity between the surrogate model and the target model with the same architecture, which might not always hold in practical applications. (2) Even if two models share the same architecture, differences in their training data or hyperparameter settings can lead to variations in the transferability of AEs~\cite{wu2018understanding}, which have not been considered in the proposed metric. (3) Note that AWT has introduced extra computational cost with two backward for calculating parameter gradients during each iteration while different samples do need different surrogate parameters to achieve better optimal. In practice, attackers should balance the trade-off between computational cost and effectiveness.

Despite the above limitations, the proposed metric serves as a valuable starting point for evaluating the transferability of AEs. It can be used as a guiding tool to identify potentially effective attacks and to inform the selection of appropriate surrogate models. Besides, further research is needed to develop more comprehensive metrics that can account for the complexities involved in real-world scenarios.

\section{Conclusion}
In this work, we focus on transferable adversarial attacks, which pose significant threats to the robustness and security of DNNs in real-world applications. Motivated by the empirical evidence that flat maxima enhance the transferability of AEs, we propose the AWT method, a novel approach aimed at finding flat adversarial regions through a max-min bi-level optimization problem. AWT leverages the interplay between the optimization of adversarial perturbations and the adjustment of the surrogate model's parameters without the requirement for extra data to achieve better transferability. Through extensive experiments on a variety of models with different architectures, we show that AWT generates AEs in flatter regions, resulting in improved transferability compared to SOTA methods. Our theoretical analysis reveals the inherent factors of the surrogate model and AEs that contribute to the transferability of AEs, and we propose a unified metric to measure the transferability of AEs across models. The enhanced transferability of AEs produced by AWT highlights the need for robust defenses and opens up new avenues for research in adversarial robustness.

\section{Acknowledgement}
This work was partly supported by the National Key Research and Development Program of China under No. 2022YFB3102100, NSFC under No. U244120033, U24A20336, 62172243 and 62402425, and the Zhejiang Provincial Natural Science Foundation under No. LD24F020002.

\bibliography{aaai25}

\newpage

\appendix

In this supplementary material, we provide contents that were not included in our main paper due to space limitations. This includes additional analysis details, additional quantitative results and proof of the proposition.

\section{Proof of the Proposition}

\setcounter{proposition}{0}

\begin{proposition}
\label{pro1}
Given a model $f_{\theta}$ with parameters $\theta$, and a perturbation $\eta$ such that $\|\eta\|_{p} \leq \kappa$ ($\kappa \rightarrow 0$), we want to prove that $\forall x\in\mathcal{D}$ and $\gamma \rightarrow 0$, there exists an perturbation $\delta$ such that:
\begin{equation}
\|f_{\theta+\eta}(x) - f_{\theta}(x+\delta)\|_{p} \leq \gamma.
\end{equation}
\end{proposition}
\textbf{Proof of the Proposition~\ref{pro1}}: Given a model $f_{\theta}$ with parameters $\theta$, and a perturbation $\eta$ such that $\|\eta\|_{p} \leq \kappa$ ($\kappa \rightarrow 0$), we want to prove that $\forall x\in\mathcal{D}$ and $\gamma \rightarrow 0$, there exists an perturbation $\delta$ such that:
\begin{equation}
\|f_{\theta+\eta}(x) - f_{\theta}(x+\delta)\|_{p} \leq \gamma.
\end{equation}

\begin{assumption}
\label{asmp1}
Continuity of $f$ with respect to $\theta$. Given the continuity of $f_{\theta}(x)$ with respect to $\theta$, for any small perturbation $\eta$ with $\|\eta\|_{p} \leq \kappa$, and $\epsilon \rightarrow 0$ as $\kappa \rightarrow 0$, there exists a small $\epsilon > 0$ such that:
\begin{equation}
\|f_{\theta+\eta}(x) - f_{\theta}(x)\|_{p} \leq \epsilon.
\end{equation}

\end{assumption}
\begin{assumption}
\label{asmp2}
Lipschitz Continuity of $f$ with respect to $x$. Given the Lipschitz continuity of $f_{\theta}(x)$ with respect to $x$, for any $\delta$, we have:
\begin{equation}
\|f_{\theta}(x + \delta) - f_{\theta}(x)\|_{p} \leq L \|\delta\|_{p}
\end{equation}
\end{assumption}
For a given $\eta$ such that $\|\eta\|_{p} \leq \kappa$, choose $\delta$ such that: $\|\delta\|_{p} \leq \frac{\epsilon}{L}$. Consider the difference:
\begin{equation}
\|f_{\theta+\eta}(x) - f_{\theta}(x+\delta)\|_{p}
\end{equation}
This can be decomposed as:
\begin{equation}
\begin{split}
    &\|f_{\theta+\eta}(x) - f_{\theta}(x+\delta)\|_{p} \leq \\
    &\|f_{\theta+\eta}(x) - f_{\theta}(x)\|_{p} + \|f_{\theta}(x) - f_{\theta}(x+\delta)\|_{p}
\end{split}
\end{equation}
Using Assumption \ref{asmp1} and \ref{asmp2}, we have:
\begin{equation}
\|f_{\theta+\eta}(x) - f_{\theta}(x)\|_{p} \leq \epsilon,
\end{equation}
and based on the inequality above, we have:
\begin{equation}
\|f_{\theta}(x) - f_{\theta}(x+\delta)\|_{p} \leq L \|\delta\|_{p} \leq L \frac{\epsilon}{L} = \epsilon
\end{equation}
Therefore,
\begin{equation}
\|f_{\theta+\eta}(x) - f_{\theta}(x+\delta)\|_{p} \leq \epsilon + \epsilon = 2\epsilon,
\end{equation}
since $\epsilon$ can be made arbitrarily small as $\kappa \rightarrow 0$, we can choose an appropriate small $\epsilon$ and corresponding $\delta$ such that $2\epsilon \leq \gamma$, and hence $\gamma \rightarrow 0$.

\begin{proposition}
\label{pro2}
    For a model $f_{\theta}$ parameterized by $\theta$ and a loss function $\ell(x, y; \theta)$, there is a positive correlation between the $L_2$ norm of the input gradient $\|\nabla_{x}\ell(x,y;\theta)\|_2$ and that of the parameter gradient $\|\nabla_{\theta}\ell(x,y;\theta)\|_2$.
\end{proposition}
\textbf{Proof of the Proposition~\ref{pro2}}: With the results in Proposition~\ref{pro1}, we start by applying the first-order Taylor expansion to both $f_{\theta+\eta}(x)$ and $f_{\theta}(x+\delta)$ around the point $(\theta, x)$:
\begin{align}
f_{\theta+\eta}(x) &\approx f_{\theta}(x) + \eta^T \nabla_{\theta} f_{\theta}(x) \\
f_{\theta}(x+\delta) &\approx f_{\theta}(x) + \delta^T \nabla_{x} f_{\theta}(x)
\end{align}
Substituting these approximations into the inequality from the Proposition~\ref{pro1}, we obtain:
\begin{equation}
\|f_{\theta}(x) + \eta^T \nabla_{\theta} f_{\theta}(x) - (f_{\theta}(x) + \delta^T \nabla_{x} f_{\theta}(x))\|_p \leq \gamma
\end{equation}

Simplifying, we get:
\begin{equation}
\|\eta^T \nabla_{\theta} f_{\theta}(x) - \delta^T \nabla_{x} f_{\theta}(x)\|_p \leq \gamma
\end{equation}

As $\gamma \rightarrow 0$ and assuming $\|\eta\|_p \rightarrow 0$, the higher-order terms in the Taylor expansions become negligible, leading to the approximation:
\begin{equation}
\eta^T \nabla_{\theta} f_{\theta}(x) = \delta^T \nabla_{x} f_{\theta}(x) + \gamma
\end{equation}
The gradients of the cross-entropy loss function $\ell(x, y; \theta)$ are defined as:
\begin{equation}
\nabla_{x}\ell(x, y; \theta) = \frac{\partial \ell(x, y; \theta)}{\partial x},
\end{equation}
\begin{equation}
\nabla_{\theta}\ell(x, y; \theta) = \frac{\partial \ell(x, y; \theta)}{\partial \theta}.
\end{equation}
Using the chain rule, we can relate the gradients of the loss function to the gradients of the model outputs:
\begin{equation}
\nabla_{x}\ell(x, y; \theta) = \frac{\partial \ell}{\partial f_{\theta}} \cdot \nabla_{x} f_{\theta}(x),
\end{equation}
\begin{equation}
\nabla_{\theta}\ell(x, y; \theta) = \frac{\partial \ell}{\partial f_{\theta}} \cdot \nabla_{\theta} f_{\theta}(x),
\end{equation}
where $\frac{\partial \ell}{\partial f_{\theta}}$ is the gradient of the loss function with respect to the model outputs $f_{\theta}(x)$.
From the previous analysis, we have shown that for small parameter perturbations $\eta$ and input perturbations $\delta$, the following approximation holds:
\begin{equation}
\eta^T \nabla_{\theta}\ell(x, y; \theta) \approx \delta^T \nabla_{x}\ell(x, y; \theta).
\end{equation}

To show the positive correlation between the L2 norms of the gradients, we consider the inner products of the gradients with $\eta$ and $\delta$:
\begin{equation}
\eta^T \nabla_{\theta}\ell(x, y; \theta) \approx \delta^T \nabla_{x}\ell(x, y; \theta).
\end{equation}
Taking the absolute value of both sides and dividing by $\|\eta\|_2$ and $\|\delta\|_2$, respectively, we obtain:
\begin{equation}
\left| \frac{\eta^T \nabla_{\theta}\ell(x, y; \theta)}{\|\eta\|_2} \right| \approx \left| \frac{\delta^T \nabla_{x}\ell(x, y; \theta)}{\|\delta\|_2} \right|.
\end{equation}
This can be further simplified to:
\begin{equation}
\left| \frac{\eta^T}{\|\eta\|_2} \right| \|\nabla_{\theta}\ell(x, y; \theta)\|_2 \approx \left| \frac{\delta^T}{\|\delta\|_2} \right| \|\nabla_{x}\ell(x, y; \theta)\|_2.
\end{equation}
Since $\eta$ and $\delta$ are small perturbations, the unit vectors $\frac{\eta}{\|\eta\|_2}$ and $\frac{\delta}{\|\delta\|_2}$ represent the directions of the perturbations. The absolute values of the inner products represent the projections of the gradients onto these directions. Given the conditions in assumptions, we conclude that the L2 norms of the input gradient $\|\nabla_{x}\ell(x,y;\theta)\|_2$ and the parameter gradient $\|\nabla_{\theta}\ell(x,y;\theta)\|_2$ are positively correlated. This is because the direction and magnitude of the gradients are related through the perturbations $\eta$ and $\delta$.

\section{Additional Analysis Details}  
\begin{figure*}[t]
    \centering
    \includegraphics[width=0.32\linewidth]{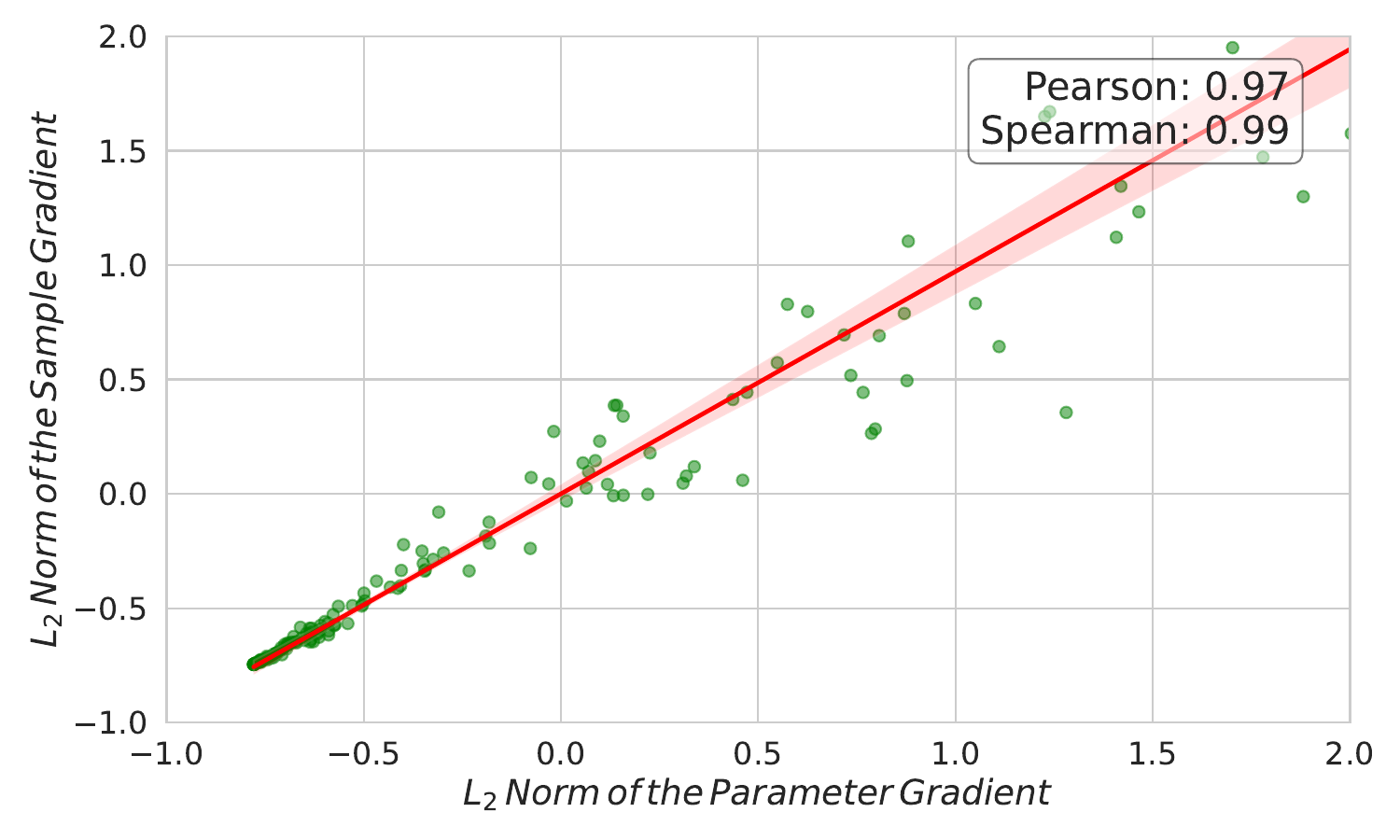}
    \includegraphics[width=0.32\linewidth]{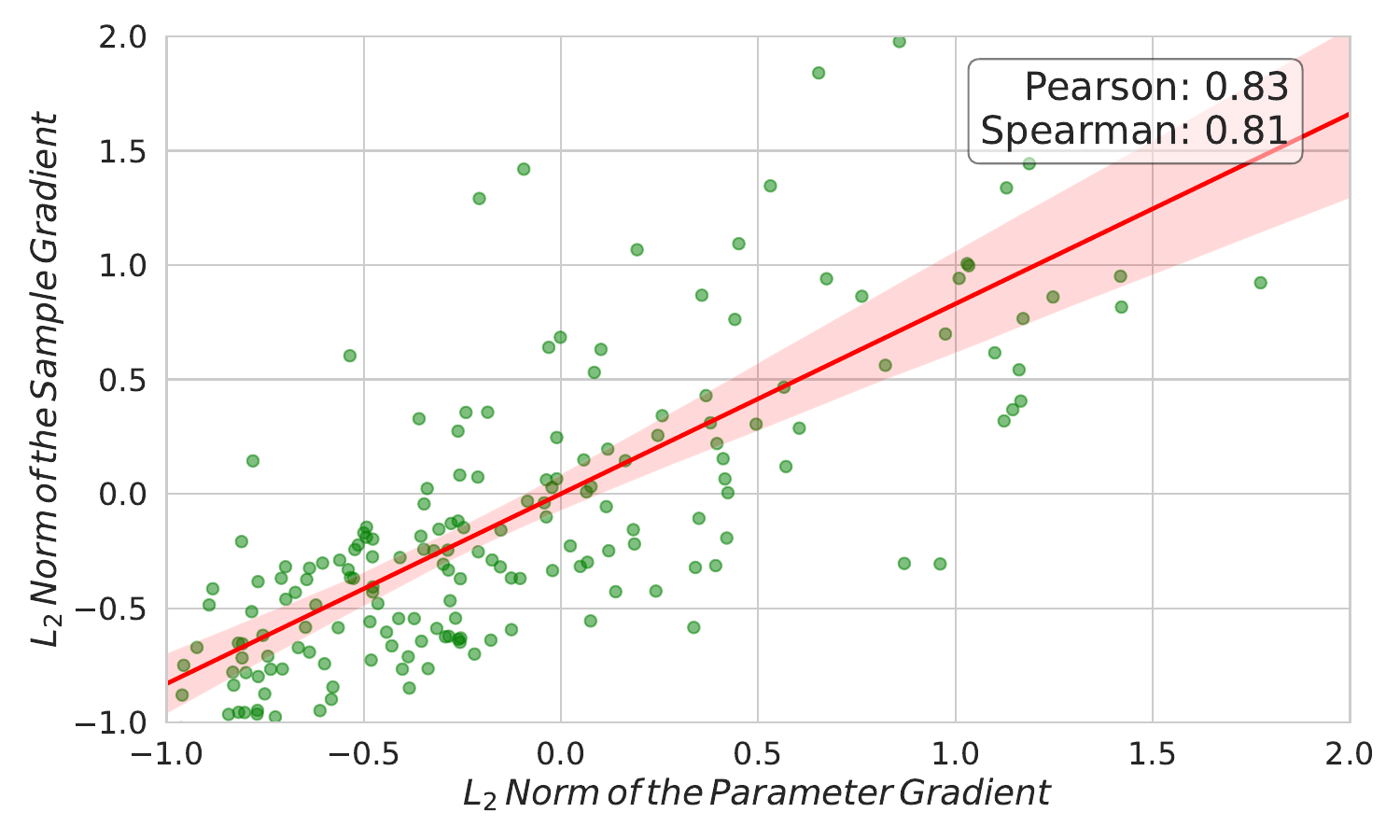}
    \includegraphics[width=0.32\linewidth]{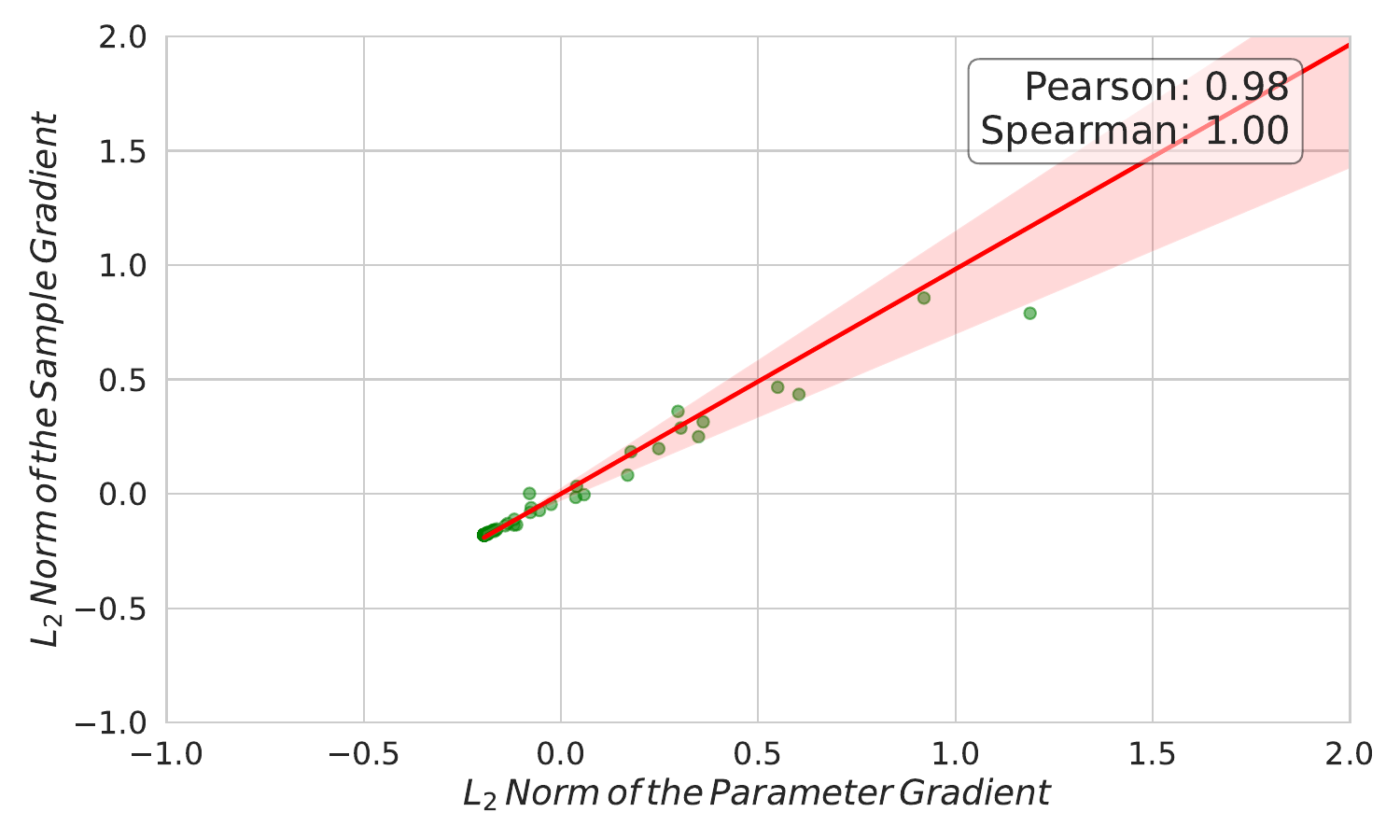}
    \includegraphics[width=0.32\linewidth]{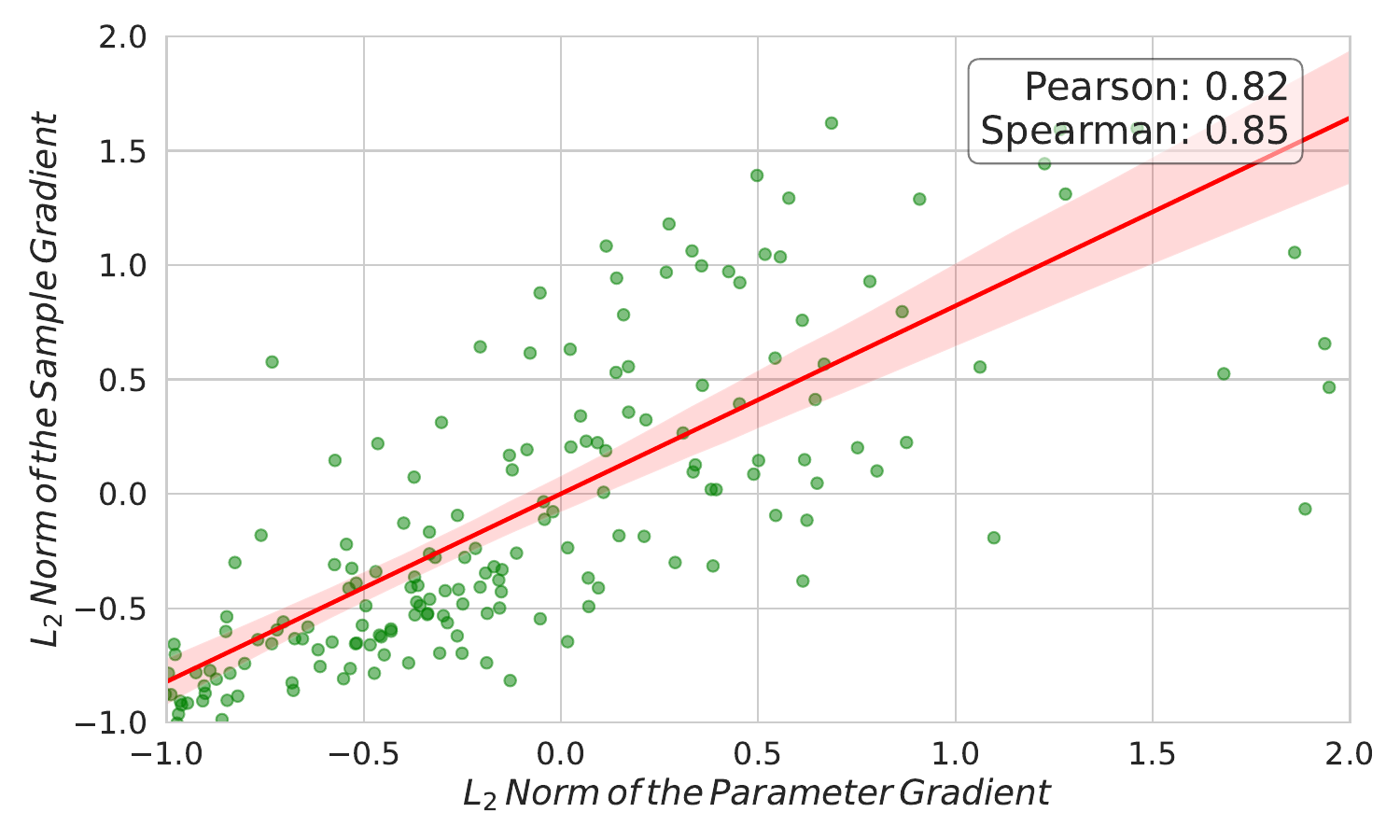}
    \includegraphics[width=0.32\linewidth]{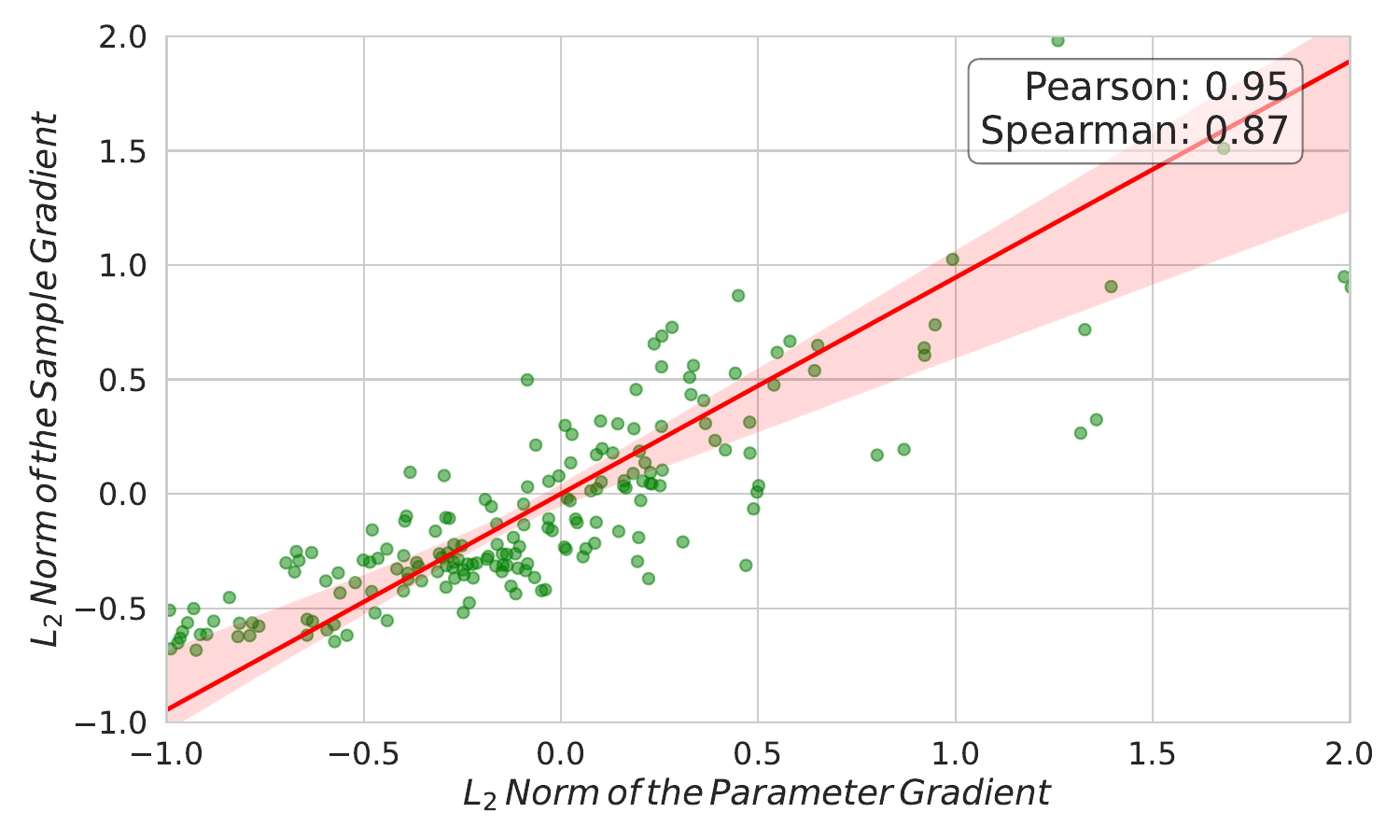}
    \includegraphics[width=0.32\linewidth]{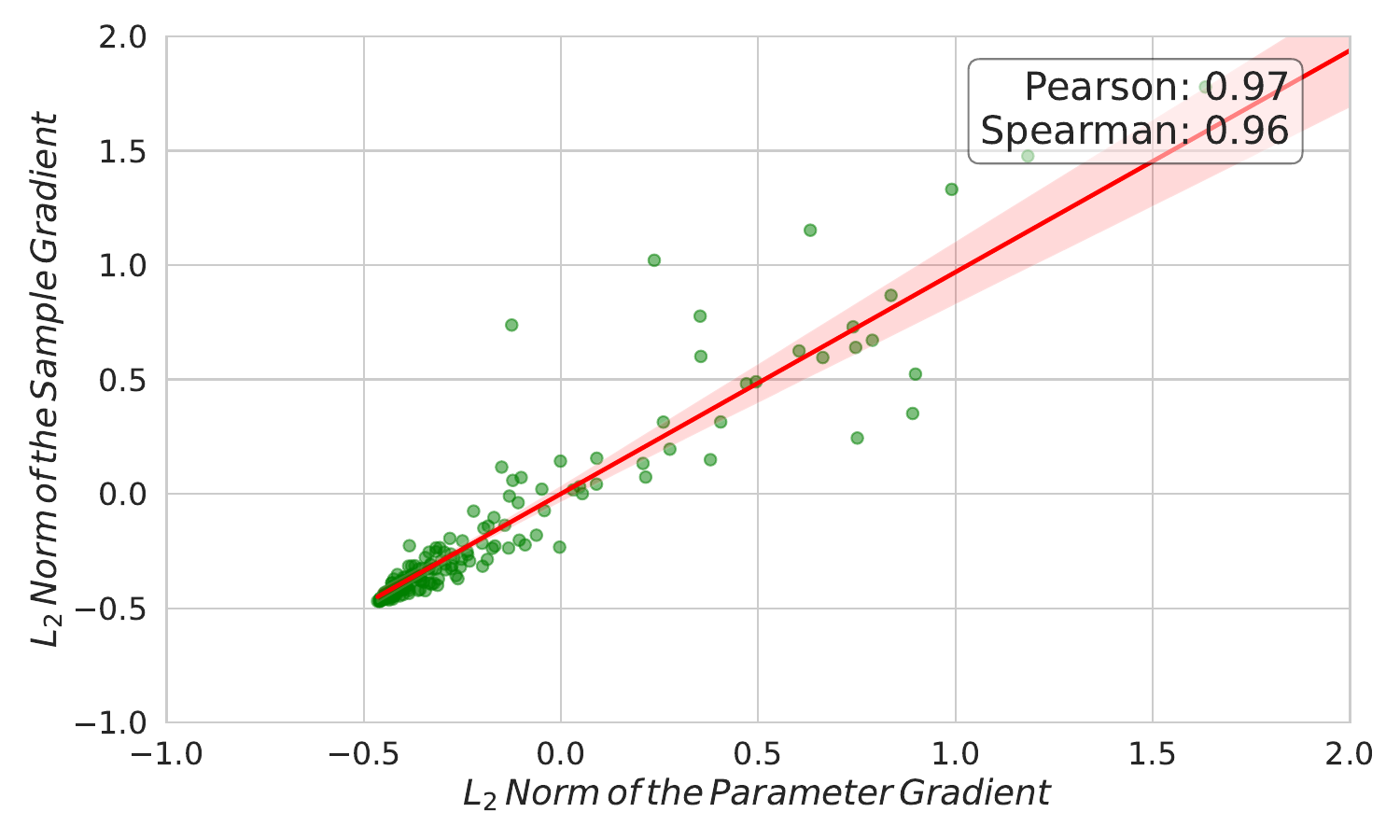}
    \caption{Correlation between the $L_2$ norms of sample and parameter gradient (with normalization: $g=\frac{g-\mu}{\sigma}$). The first row includes models: VGG16, ResNet50 and InceptionV4. The second row includes models: Swin-s, ConVit and DeiT-B. }
    \label{fig:gradient_correlation}
\end{figure*}

In this section, we provide additional analysis details that were not included in the main paper due to space limitations. Specifically, we present detailed visualizations and discussions regarding the correlation between the $L_2$ norms of the input gradient and the parameter gradient for different models. We examine the correlation between the $L_2$ norms of the input gradient $\|\nabla_{x}\ell(x,y;\theta)\|_2$ and the parameter gradient $\|\nabla_{\theta}\ell(x,y;\theta)\|_2$ for a variety of models. To normalize the gradients and ensure comparability across different models, where we also apply the following normalization to each gradient vector:
\begin{equation}
g = \frac{g - \mu}{\sigma},
\end{equation}
where $g$ is the gradient vector, $\mu$ is the mean of the gradient values, and $\sigma$ is the standard deviation of the gradient values. Figure~\ref{fig:gradient_correlation} shows scatter plots for six different models, including three CNN-based models (VGG16, ResNet50, InceptionV4) and three Transformer-based models (Swin-S, ConViT, DeiT-B). Each point in the scatter plot represents a pair of normalized $L_2$ norms of the input and parameter gradients for a single input sample. We observe a positive correlation between the two types of gradients for all models, indicating that the magnitude of the input gradient is indicative of the magnitude of the parameter gradient. The positive correlation observed in Figure~\ref{fig:gradient_correlation} suggests that the magnitude of the input gradient can serve as a proxy for the magnitude of the parameter gradient. This finding supports the theoretical underpinning of our approach, where we use the input gradient to infer properties about the parameter gradient, which is crucial for the development of our adversarial attack algorithm.

\section{Additional Quantitative Results}

In this section, we provide additional quantitative results that complement the findings presented in the main paper. These results include a more extensive comparison with baseline methods of the performance of our proposed AWT method across different models.

\subsection{Comparison with Baseline Methods}

Tables~\ref{tab:full_transformer_attack} and~\ref{tab:full_cnn_attack} present the ASRs for a wider range of models, including both CNN- and Transformer-based models, using our proposed AWT method and several baseline methods.

\subsubsection{CNN-Based Models}

\begin{table*}[t]
\centering
\tabcolsep=0.15cm
\renewcommand{\arraystretch}{1.2}
\begin{tabular}{|c|c|cccccccc|}
\hline
Model & Attack & Inc-v4 & Inc-Res-v2 & Inc-v3 & RN18 & RN50 & RN101 & VGG-13 & VGG-16 \\ \hline
\multirow{9}{*}{RN50} & MI & 26.7±0.2 & 21.3±0.5 & 32.8±0.8 & 39.7±0.6 & \textit{79.9±0.4} & 30.1±0.2 & 45.3±0.7 & 40.4±0.5 \\
 & NI & 26.3±0.5 & 21.1±0.7 & 32.8±0.2 & 41.4±0.5 & \textit{79.0±0.5} & 29.3±0.5 & 47.3±0.5 & 41.8±0.3 \\
 & VMI & 46.9±0.5 & 40.1±0.6 & 49.1±0.3 & 55.1±0.5 & \textit{88.6±0.2} & 54.1±0.6 & 58.9±0.5 & 56.5±0.4 \\
 & VNI & 45.5±0.3 & 38.0±0.9 & 47.6±0.6 & 56.5±0.3 & \textit{93.9±0.2} & 54.8±0.7 & 61.1±0.6 & 58.2±0.8 \\
 & EMI & 43.1±0.7 & 35.0±0.5 & 48.1±0.9 & 63.5±1.0 & \textit{97.2±0.1} & 51.5±0.6 & 67.5±0.8 & 63.9±0.8 \\
 & RAP & 25.1±0.2 & 20.1±0.2 & 30.8±0.7 & 37.7±0.7 & \textit{76.7±0.7} & 27.4±0.3 & 43.5±0.7 & 38.9±0.5 \\
 & PGN & \ul{73.2±0.3} & \ul{69.5±0.3} & \ul{75.3±0.3} & \ul{80.7±0.2} & \ul{\textit{95.3±0.1}} & \ul{78.8±0.2} & \ul{80.4±0.2} & \ul{80.2±0.2} \\
 & NCS & 67.8±0.1 & 63.9±0.2 & 68.5±0.2 & 74.5±0.4 & \textit{90.2±0.0} & 72.5±0.2 & 75.9±0.2 & 73.9±0.1 \\
 & AWT & \textbf{78.0±0.3} & \textbf{74.6±0.5} & \textbf{79.4±0.3} & \textbf{85.8±0.2} & \textit{\textbf{96.4±0.1}} & \textbf{83.0±0.3} & \textbf{85.0±0.1} & \textbf{85.4±0.3} \\ \hline
\multirow{9}{*}{RN101} & MI & 36.6±0.3 & 30.3±0.1 & 40.5±0.6 & 48.9±0.6 & 47.9±1.0 & \textit{86.0±0.3} & 50.4±0.3 & 46.9±0.4 \\
 & NI & 36.7±0.7 & 29.9±0.5 & 40.2±0.4 & 50.0±0.7 & 48.4±0.2 & \textit{86.7±0.5} & 52.3±0.3 & 48.0±0.9 \\
 & VMI & 55.2±0.3 & 50.2±0.4 & 57.2±0.4 & 61.9±0.6 & 65.8±0.5 & \textit{89.2±0.2} & 62.6±0.5 & 60.0±0.2 \\
 & VNI & 55.4±0.5 & 47.9±0.6 & 56.2±1.0 & 64.5±0.7 & 70.5±0.3 & \textit{96.8±0.2} & 64.6±0.7 & 61.7±0.3 \\
 & EMI & 59.6±0.8 & 51.5±1.1 & 61.3±0.4 & 73.6±0.4 & 77.7±0.4 & \textit{98.2±0.1} & 71.9±0.5 & 70.1±0.4 \\
 & RAP & 34.2±0.4 & 27.6±1.3 & 38.4±0.9 & 47.2±0.7 & 43.8±0.9 & \textit{80.8±0.5} & 48.9±0.3 & 44.9±0.7 \\
 & PGN & \ul{82.1±0.1} & \ul{78.3±0.2} & \ul{81.0±0.2} & \ul{84.1±0.1} & \ul{86.7±0.2} & \ul{\textit{96.0±0.1}} & \ul{82.5±0.2} & \ul{82.1±0.3} \\
 & NCS & 75.0±0.4 & 69.7±0.1 & 74.2±0.2 & 78.6±0.2 & 79.8±0.2 & \textit{90.5±0.2} & 78.0±0.2 & 76.5±0.3 \\
 & AWT & \textbf{83.0±0.1} & \textbf{79.8±0.2} & \textbf{82.5±0.1} & \textbf{86.1±0.2} & \textbf{88.0±0.1} & \textit{\textbf{96.2±0.1}} & \textbf{84.8±0.1} & \textbf{84.8±0.3} \\ \hline
\multirow{9}{*}{Inc-v3} & MI & 45.7±1.4 & 40.6±0.9 & 98.3±0.0 & 42.3±0.9 & 24.9±0.4 & 22.6±0.4 & 44.8±1.0 & 39.5±1.3 \\
 & NI & 50.8±1.3 & 45.8±0.6 & 98.5±0.0 & 47.5±0.9 & 28.6±0.8 & 25.7±1.0 & 48.7±0.4 & 44.0±0.4 \\
 & VMI & 63.6±1.0 & 59.9±0.4 & 98.8±0.1 & 54.3±0.7 & 37.6±0.5 & 36.4±0.3 & 55.3±1.1 & 50.9±0.8 \\
 & VNI & 61.7±0.2 & 58.2±0.6 & 99.5±0.0 & 53.7±0.8 & 35.3±0.6 & 33.4±0.6 & 53.3±0.5 & 50.2±0.6 \\
 & EMI & 61.1±0.2 & 57.0±0.6 & 99.9±0.0 & 58.1±0.8 & 34.8±0.2 & 31.9±0.4 & 56.2±0.3 & 52.2±0.9 \\
 & RAP & 48.0±0.6 & 42.5±0.9 & 99.7±0.0 & 46.9±0.7 & 28.1±0.6 & 24.2±0.4 & 48.3±1.0 & 44.2±1.1 \\
 & PGN & 80.0±0.6 & 77.8±0.8 & 100.0±0.0 & 71.3±0.8 & 49.4±0.6 & 47.3±0.3 & 66.8±0.5 & 66.5±0.4 \\
 & NCS & \ul{83.7±0.1} & \ul{82.5±0.2} & \ul{99.3±0.0} & \ul{75.9±0.4} & \ul{57.5±0.3} & \textbf{56.8±0.5} & \ul{72.3±0.2} & \ul{71.6±0.7} \\
 & AWT & \textbf{85.1±0.8} & \textbf{83.2±0.6} & \textbf{99.8±0.0} & \textbf{79.6±0.7} & \textbf{57.9±0.4} & \ul{55.3±0.4} & \textbf{75.3±0.5} & \textbf{74.8±0.8} \\ \hline
\end{tabular}
\caption{(Complete Table) The attack success rates (\%±std, over 10 random runs) of nine gradient-based attacks on four normally trained CNN-based models. The results of White-box attacks are shown with \textit{italic}. The best and second results are \textbf{bold} and \ul{underline}, respectively.}
\label{tab:full_cnn_attack}
\end{table*}

Table~\ref{tab:full_cnn_attack} shows the ASRs for a variety of CNN-based models, including ResNet50, ResNet101, and InceptionV3. Across these models, our proposed AWT method consistently outperforms other attacks, achieving the highest attack success rates. PGN and NCS are often the closest competitors to AWT, ranking second in many cases. The attack success rates vary significantly depending on the model architecture, with AWT showing a clear improvement over other methods, especially on complex models like ResNet-50 and ResNet-101.

\subsubsection{Transformer-Based Models}
\begin{table*}[]
\centering
\tabcolsep=0.15cm
\renewcommand{\arraystretch}{1.3}
\begin{tabular}{|c|c|ccccccccc|}
\hline
Model & Attack & ViT-B/16 & ViT-B/32 & Swin-T & Swin-S & ConViT & LeViT & CAiT-S & DeiT-B & PiT-B \\ \hline
\multirow{9}{*}{RN50} & MI & 11.7±0.3 & 11.6±0.2 & 25.1±0.6 & 20.6±0.5 & 11.5±0.3 & 17.8±0.2 & 12.3±0.4 & 10.9±0.5 & 15.5±0.3 \\
 & NI & 11.3±0.3 & 11.3±0.3 & 25.2±0.1 & 20.9±0.4 & 11.2±0.2 & 17.7±0.2 & 11.6±0.4 & 10.7±0.2 & 14.5±0.1 \\
 & VMI & 28.6±0.2 & 21.6±0.3 & 42.5±0.3 & 38.1±0.3 & 27.4±0.4 & 36.0±0.4 & 29.4±0.4 & 28.5±0.9 & 33.3±0.3 \\
 & VNI & 23.0±0.7 & 17.4±0.4 & 40.8±0.5 & 35.4±0.3 & 23.0±0.2 & 33.5±0.6 & 24.4±0.4 & 23.6±0.3 & 29.3±0.4 \\
 & EMI & 19.3±0.2 & 15.6±0.3 & 39.5±0.6 & 32.8±0.1 & 19.8±0.4 & 30.6±0.4 & 20.9±0.5 & 19.2±0.4 & 25.1±0.3 \\
 & RAP & 10.8±0.2 & 10.9±0.2 & 23.9±0.1 & 19.9±0.6 & 10.6±0.1 & 16.5±0.2 & 11.4±0.1 & 9.9±0.2 & 14.6±0.4 \\
 & PGN & \ul{49.2±0.4} & \ul{40.8±0.2} & \ul{66.0±0.1} & \ul{59.2±0.2} & \ul{46.8±0.4} & \ul{60.4±0.6} & \ul{50.9±0.2} & \ul{49.5±0.3} & \ul{53.0±0.2} \\
 & NCS & 49.1±0.3 & 39.7±0.3 & 63.7±0.1 & 59.4±0.2 & 48.0±0.2 & 58.8±0.5 & 51.5±0.2 & 50.6±0.3 & 54.1±0.5 \\
 & AWT & \textbf{59.5±0.4} & \textbf{47.3±0.4} & \textbf{73.0±0.4} & \textbf{68.1±0.4} & \textbf{57.3±0.9} & \textbf{69.7±0.4} & \textbf{61.6±0.4} & \textbf{60.5±0.2} & \textbf{63.5±0.4} \\
 \hline
\multirow{9}{*}{RN101} & MI & 18.5±0.4 & 14.4±0.2 & 31.6±0.6 & 28.1±0.4 & 18.7±0.5 & 25.8±0.9 & 19.2±0.1 & 18.3±0.3 & 23.3±0.4 \\
 & NI & 17.3±0.3 & 13.5±0.3 & 31.2±0.6 & 27.8±0.9 & 17.8±0.5 & 25.5±0.2 & 18.1±0.4 & 16.9±0.2 & 22.0±0.2 \\
 & VMI & 36.0±0.2 & 26.2±0.2 & 48.8±0.3 & 45.3±0.5 & 35.8±0.6 & 44.3±0.7 & 37.9±0.4 & 37.0±0.6 & 41.0±0.9 \\
 & VNI & 31.0±0.8 & 21.9±0.7 & 47.5±0.4 & 42.2±1.1 & 31.0±0.6 & 42.1±0.9 & 33.5±0.5 & 32.2±0.7 & 37.8±1.1 \\
 & EMI & 30.4±1.0 & 21.3±0.2 & 51.5±0.5 & 44.8±0.5 & 30.8±0.4 & 44.8±0.6 & 33.5±0.5 & 31.8±0.4 & 37.7±0.5 \\
 & RAP & 16.4±0.3 & 13.6±0.1 & 29.4±0.3 & 25.6±0.3 & 16.7±0.2 & 23.8±0.1 & 17.3±0.3 & 15.6±0.2 & 20.9±0.2 \\
 & PGN & \ul{63.0±0.5} & \ul{51.9±0.3} & \ul{73.4±0.3} & \ul{69.4±0.2} & \ul{60.2±0.5} & \ul{70.7±0.4} & \ul{65.9±0.3} & \ul{63.5±0.1} & \ul{65.8±0.1} \\
 & NCS & 58.5±0.1 & 48.3±0.4 & 69.7±0.2 & 66.3±0.2 & 57.5±0.3 & 66.0±0.3 & 62.5±0.2 & 59.6±0.1 & 62.6±0.1 \\
 & AWT & \textbf{67.8±0.2} & \textbf{55.2±0.2} & \textbf{77.2±0.4} & \textbf{73.8±0.2} & \textbf{66.1±0.4} & \textbf{75.5±0.2} & \textbf{71.7±0.1} & \textbf{68.9±0.2} & \textbf{70.3±0.4} \\ \hline
\multirow{9}{*}{Inc-v3} & MI & 13.4±0.2 & 13.9±0.3 & 23.5±0.5 & 20.8±0.2 & 13.8±0.2 & 20.6±0.5 & 13.6±0.4 & 11.7±0.1 & 14.7±0.3 \\
 & NI & 14.4±0.5 & 14.4±0.3 & 25.7±0.6 & 22.9±0.1 & 15.3±0.7 & 23.6±0.3 & 15.5±0.3 & 13.3±0.3 & 15.6±0.4 \\
 & VMI & 22.2±0.3 & 19.9±0.4 & 33.3±0.3 & 30.4±0.5 & 22.3±0.2 & 32.2±0.3 & 23.8±0.4 & 21.1±0.4 & 23.4±0.9 \\
 & VNI & 19.5±0.3 & 17.7±0.4 & 32.1±0.4 & 28.2±0.8 & 20.4±0.3 & 29.6±0.6 & 21.1±0.6 & 18.9±0.3 & 20.6±0.3 \\
 & EMI & 19.2±0.4 & 18.0±0.5 & 31.6±0.6 & 27.5±1.2 & 19.4±0.5 & 29.0±0.7 & 19.7±0.5 & 17.6±0.3 & 19.4±0.3 \\
 & RAP & 14.6±0.1 & 14.1±0.3 & 25.3±0.3 & 21.2±0.2 & 14.8±0.2 & 22.0±0.5 & 14.3±0.4 & 11.9±0.1 & 15.0±0.1 \\
 & PGN & 31.7±0.4 & 28.1±0.4 & 44.4±0.7 & 38.0±0.4 & 30.3±0.3 & 43.4±0.3 & 31.5±0.3 & 29.6±0.5 & 30.4±0.5 \\
 & NCS & \ul{38.2±0.4} & \ul{32.7±0.8} & \ul{52.1±0.4} & \ul{44.7±0.6} & \ul{36.8±0.2} & \ul{51.9±0.3} & \ul{39.9±0.4} & \textbf{37.4±0.2} & \ul{37.6±0.6} \\
 & AWT & \textbf{39.2±1.0} & \textbf{34.7±0.4} & \textbf{52.5±0.6} & \textbf{45.6±0.7} & \textbf{37.7±0.2} & \textbf{52.7±1.1} & \textbf{40.3±1.1} & \ul{37.1±0.9} & \textbf{37.9±0.2} \\ \hline
\end{tabular}
\caption{(Complete Table) The attack success rates (\%±std, over 10 random runs) of nine gradient-based attacks on four normally trained Transformer-based models. The best and second results are \textbf{bold} and \ul{underline}, respectively.}
\label{tab:full_transformer_attack}
\end{table*}

Table~\ref{tab:full_transformer_attack} presents the ASRs for Transformer-based models. Similar to the CNN-based models, our AWT consistently outperforms other gradient-based attacks across different Transformer-based models, showing a clear improvement in terms of attack success rates. The results suggest that AWT is a robust and effective method for generating AEs against these models.

\subsection{Discussion}

The additional quantitative results reinforce the conclusions drawn in the main paper. Our AWT method consistently outperforms the baseline methods across both CNN-based and Transformer-based models, showcasing its robustness and effectiveness in generating transferable adversarial examples. The improvements in ASR are particularly notable for Transformer-based models, indicating the potential of our method in enhancing the transferability of adversarial examples in the context of modern deep learning architectures.

\end{document}